\documentclass[aps,reprint,superscriptaddress,nofootinbib,showpacs]{revtex4-1}
\usepackage{amsmath,latexsym,amssymb,hyperref,graphicx}
\usepackage{slashed}

\hypersetup{colorlinks,citecolor= nicegreen,linkcolor= niceblue}
\usepackage{color}
\definecolor{nicered}{rgb}{0.7,0.1,0.1}
\definecolor{nicegreen}{rgb}{0.1,0.5,0.1}
\definecolor{niceblue}{rgb}{0.1,0.1,0.8}

\newcommand{\beq}{\begin{equation}}
\newcommand{\eeq}{\end{equation}}
\newcommand{\bea}{\begin{eqnarray}}
\newcommand{\eea}{\end{eqnarray}}

\begin{document}

\title{Selfgravitating $SU(5)$ Higgs domain walls as a braneworlds}
\author{Nelson Pantoja}
\affiliation{Centro de F\'isica Fundamental, Universidad de Los Andes,
M\'erida, Venezuela}

\date{\today}

\begin{abstract}

Five-dimensional domain walls in gauged $SU(5)$ generate a position-dependent 
symmetry breaking pattern along the additional dimension. We analize the 
perturbative stability and the 4D spectrum of these walls in the self-gravitating case, 
in terms of diffeomorphism- and Lie algebra gauge-invariant field fluctuations. We 
show that tachyonic modes are absent, ensuring perturbative stability. As expected, 
gravitational tensor and vector fluctuations behave like its counterparts in the 
standard $Z_2$ domain walls. All the Lie algebra valued fluctuations exhibit towers 
of 4D massive modes which propagate in the bulk, with a continuous spectrum 
starting from zero. All the would-be 4D Nambu-Goldstone fields, which are 
gravitationally trapped in the case of a global symmetry, are non-trivially absent. 
However we find no localizable 4D gauge bosons, either massless or massive. 
Instead, quasi-localizable discrete 4D massive modes for the gauge field fluctuations 
are  found, along the spontaneously broken directions.  
\end{abstract}


\maketitle

\section{Introduction}     

It is well known that field theoretic domain walls, arising in abelian 
$Z_2$-symmetric 5D Einstein-Scalar Field theories, provide regularizations 
of the Randall-Sundrum braneworld \cite{Randall:1999vf} which preserve 4D gravity 
on the core of the wall. However, it is perhaps not so familiar 
the fact that it is also possible to consider domain walls, 
generated by a scalar field transforming non-trivially under a non-abelian group, that 
break a continuous internal symmetry in addition to the $Z_2$ symmetry.

In flat space $SU(5) \times Z_2$ theories with a single adjoint scalar 
$\boldsymbol{\Phi}$ and symmetry breaking 
$SU(5) \times Z_2\rightarrow H={SU(3)\times SU(2) \times U(1)}/({Z_3\times Z_2})$, 
there exist domain walls which interpolate non-trivially between 
the two disconnected sectors of the vacuum manifold 
\cite{Vachaspati:2001pw,Pogosian:2001fm,Vachaspati:2003zp}. In these walls, the 
unbroken symmetries far away of the wall, $H_\pm$, and on its core, 
$H_0=H_+\cap H_-$, are such that $H_+$ and $H_-$, though isomorphic, are 
differently embedded in $SU(5)$. Non-abelian domain 
walls of this sort are very interesting by themselves as well as in connection to 
the solitonic nature of fundamental branes  \cite{Dvali:1996xe,Dvali:2002fi}. 
 
Non-abelian domain walls (rather their extensions to the gravitating case), may be 
relevant within the context of braneworlds. In this direction, the idea of a 
braneworld generated by a domain wall that breaks a gauge symmetry group $G$ in 
addition to the $Z_2$ discrete symmetry was put forward in 
Ref.\cite{Davidson:2002eu}. Explicit flat space-time 
realizations has been discussed for a $O(10)$ symmetry in Ref.\cite{Shin:2003xy} 
and, assuming gauge field localization \textit{via} the Dvali-Shifman mechanism 
\cite{Dvali:1996xe}, for a $E_6$ invariant theory in Ref.\cite{Davidson:2007cf}. 
The last reference gave also a treatment for dynamical 
localization of fermions in the model.

Non-abelian domain walls in theories with gravity have been also considered within 
the braneworld context. It has been shown \cite{Melfo:2011ev} that domain wall 
configurations $(\boldsymbol{\Phi}^k; g^k_{ab})$ in global $SU(5)\times Z_2$ 
Einstein-Scalar Field 5D theories exist, in which the curvature of the metric 
$g^k_{ab}$ is a regularization of the curvature of the Randall-Sundrum brane. 
Analysis of the diffeomorphism-invariant fluctuations of these systems 
reveals, besides its perturbative stability, an interesting gravitationally trapped 
content from the point of view of 4D observers \cite{Pantoja:2015yin}. In particular, 
there are as many normalizable 4D massless scalar modes as there are broken 
generators (i.e. that do not commute with $\boldsymbol{\Phi}^k$). Since the domain 
wall configuration preserve $H_0$ as the largest global Lie algebra symmetry, the 
above gravitationally localized 4D massless scalar modes may be identified as (the 
4D zero modes of) the Nambu-Goldstone fields associated to the partial breaking 
$SU(5)\times Z_2\rightarrow H_0$ \cite{Pantoja:2015yin}. 

Similar results to the above ones have 
been found in Ref.\cite{Chavez:2016sbv} for the selfgravitating versions of the flat 
space-time $O(10)$ domain wall braneworlds of Ref.\cite{Shin:2003xy}. Thus, in 
domain wall braneworlds where the domain wall is used to model spontaneous 
symmetry breaking of continuous global symmetries, the inclusion of gravity leads to 
massless scalars localized on its core as the Nambu-Goldstone bosons associated to 
the broken symmetries. 

In view of the results of Refs.\cite{Pantoja:2015yin,Chavez:2016sbv} and being the 
analysis at the reach of a classical perturbative treatment, for gauge 
couplings sufficiently small, the obvious next step is to look for the fate of the 
gravitationally trapped 4D Nambu-Goldstone bosons and the behavior of the gauge 
fields in 5D selfgravitating Higgs domain walls as braneworls.
 
In order to carry out the above program, in Sec.\ref{SU_5_branes} we obtain the 
gauged versions of the selfgravitating global $SU(5)$ domain walls of 
Ref.\cite{Melfo:2011ev}. Next, in Sec.\ref{fluctuations}, after a brief discussion on 
linear perturbations of the Einstein-Yang-Mills-Higgs system and their behavior 
under diffeomorphisms and Lie algebra gauge transformations, the 
linearized field equations for the chosen set of diffeomorphism- and 
Lie algebra gauge-invariant fluctuations around the domain wall backgrounds 
of Sec.\ref{SU_5_branes} are derived. 

The dimensional reduction and the analysis of the 4D modes is carried out in 
Sec.\ref{4D_modes}. There we show the absence of tachyonic modes for all these 
gauge-invariant fluctuations and hence the perturbative stability of the domain wall 
configurations considered. We show the absence of localizable 4D massless scalar 
modes. No localizable 4D massless nor 4D massive modes for the gauge field 
fluctuations are found. We show the existence of quasi-localizable discrete 
4D massive modes for the gauge field fluctuations along the spontaneously broken 
gauge sectors. Related issues to the gauge fixing approach are also discussed. A 
summary and conclusions are given in Sec.\ref{conclusions}.
  
\section{Self-gravitating local $SU(5)$ domain walls}\label{SU_5_branes}

Let us consider the 5D theory\footnote{We use units were $G = c = 1$}  
\begin{eqnarray}\label{theory1}
S=\int d^4x\,dy\, \sqrt{-g}&&\left[\frac{1}{2}R-g^{ab}\text{Tr}\{\mathbf{D}_a{\boldsymbol{\Phi}} \mathbf{D}_b{\boldsymbol{\Phi}}\} - V({\boldsymbol{\Phi}})\nonumber\right.\\ &&\,\,\,\,\left.-\frac{1}{2}g^{ac}g^{bd}\text{Tr}\{\mathbf{F}_{ab}\mathbf{F}_{cd}\}\right],
\end{eqnarray}
where $R$ is the scalar curvature of the metric $g_{ab}$, $g=\text{det}(g_{ab})$, 
${\boldsymbol{\Phi}}$ is a scalar field that transforms in the adjoint representation of 
$SU(5)$,
\begin{equation}
\mathbf{D}_a\boldsymbol{\Phi}= \nabla_a\boldsymbol{\Phi} + i\texttt{g} [\boldsymbol{A}_a,\boldsymbol{\Phi}]
\end{equation}
is the gauge covariant derivative of $\boldsymbol{\Phi}$ with $\nabla_{c}g_{ab}=0$,
 \begin{equation}\label{SFT}
 \mathbf{F}_{ab}=\nabla_a\boldsymbol{A}_b- \nabla_b\boldsymbol{A}_a + i\texttt{g}[\boldsymbol{A}_a,\boldsymbol{A}_b]
 \end{equation}
is the field strength tensor of the gauge field $\boldsymbol{A}_a$ and $V({\boldsymbol{\Phi}})$ a sixth-order potential of the form
\begin{eqnarray}\label{potential1}
\!\!\!\!V({\boldsymbol{\Phi}})&&=\!\!V_0-\mu^2 \text{Tr}\{{\boldsymbol{\Phi}}^2\}+h (\text{Tr}\{{\boldsymbol{\Phi}}^2\})^2+\lambda \text{Tr}\{{\boldsymbol{\Phi}}^4\} \nonumber\\&&\!\!+ \alpha (\text{Tr}{\boldsymbol{\Phi}}^2\})^3 + \beta (\text{Tr} \{{\boldsymbol{\Phi}}^3\})^2+ \gamma \text{Tr}\{{\boldsymbol{\Phi}}^4\}\text{Tr}\{\boldsymbol{\Phi}^2\}.
\end{eqnarray}

Besides being invariant under general space-time diffeomorphisms, the theory 
(\ref{theory1}) is invariant under local $SU(5)$ gauge transformations 
\begin{itemize}
\item {$\boldsymbol{\Phi}\mapsto \mathbf{U}\boldsymbol{\Phi} \mathbf{U}^\dagger$},
\item $\boldsymbol{A}_a\mapsto \mathbf{U}\boldsymbol{A}_a\mathbf{U}^\dagger + ({i}/{\texttt{g}})(\nabla_a \mathbf{U})\mathbf{U}^\dagger$
\item $g_{ab}\mapsto g_{ab}$,                    
\end{itemize}
where {$\mathbf{U}=\exp\{-i\sigma_q\mathbf{T}^q\}$}, with $\sigma_q=\sigma_q(x,y)$ 
finite functions on space-time such that $\mathbf{U}$ tends to the identity at spatial 
infinite and $\mathbf{T}^q, \,\,q=1,\ldots ,24$, are traceless hermitian generators of the Lie algebra $\mathfrak{su}(5)$ of $SU(5)$, normalized so that $\text{Tr}\{\mathbf{T}^q\mathbf{T}^p\}=(1/2)\delta^{qp}$. It is also invariant under 
\begin{itemize}
\item $Z_2: \boldsymbol{\Phi} \mapsto -\boldsymbol{\Phi}$, $\quad Z_2 \notin SU(5)$
\end{itemize}
which leave the gravitational and gauge field sectors invariant.

The field equations, following from (\ref{theory1}), are given by 
\begin{equation}\label{field_eq1}
 R_{ab}-\frac{1}{2}g_{ab}R=T_{ab},
\end{equation}
where
\begin{eqnarray}\label{field_eq2}
T_{ab}&=&2\text{Tr}\{\mathbf{D}_{a}\boldsymbol{\Phi} \mathbf{D}_{b}\boldsymbol{\Phi}\}-g_{ab}\left(g^{cd}\text{Tr}\{\mathbf{D}_{c}\boldsymbol{\Phi} \mathbf{D}_{d}\boldsymbol{\Phi}\}+V(\boldsymbol{\Phi})\right)\nonumber\\
&&+ 2\text{Tr}\{\mathbf{F}_{ac}\mathbf{F}_b^{\,\,\,c}\} -\frac{1}{2}g_{ab}\text{Tr}\{\mathbf{F}_{cd}\mathbf{F}^{cd}\},
\end{eqnarray}
\begin{equation}\label{field_eq3}
g^{ab}\mathbf{D}_{a}(\mathbf{D}_{b}\boldsymbol{\Phi})=\frac{\partial V(\boldsymbol{\Phi})}{\partial\phi_q}\mathbf{T}^q,\quad \boldsymbol{\Phi}=\phi_q\mathbf{T}^q
\end{equation}
and
\begin{equation}\label{field_eq4}
\mathbf{D}_a\mathbf{F}^{ab} - i\texttt{g}[\boldsymbol{\Phi}, \mathbf{D}^b\boldsymbol{\Phi}]=0.
\end{equation}

Next, assuming that the geometry preserves 4D-Poincare invariance, the 5D 
manifold is endowed with a metric of the form
\begin{equation}\label{metric-ansatz}
g_{ab}= e^{2A(y)}\eta_{\mu\nu}dx^{\mu}_adx^{\nu}_b + dy_a dy_b,
\end{equation}
with $\eta_{\mu\nu}=\text{diag}(-1,+1,+1,+1)$. Now, in the $\{x^\mu,y\}$ coordinate 
system, we seek for field configurations 
$(\tilde{\boldsymbol{\Phi}}^k,\tilde{\boldsymbol{A}}^k_a;g^k_{ab})$ such that
\begin{equation}\label{kink_gauge}
\tilde{\boldsymbol{\Phi}}^k(x,y)=\mathbf{U}\boldsymbol{\Phi}^k(y) \mathbf{U}^\dagger,
\qquad
\tilde{\boldsymbol{F}}_{ab}=0.
\end{equation}
Then $\tilde{\boldsymbol{\!A}}^k_a$ is given by a pure gauge 
\begin{equation}\label{pure_gauge}
\tilde{\boldsymbol{\!A}}^k_a(x,y)= +\frac{i}{\texttt{g}}(\partial_a\mathbf{U})\mathbf{U}^\dagger
\end{equation}
and (\ref{field_eq4}) requires
\begin{equation}\label{kink_property}
[\tilde{\boldsymbol{\Phi}}^k, \tilde{\mathbf{D}}_b\tilde{\boldsymbol{\Phi}}^k]=\mathbf{U}[\boldsymbol{\Phi}^k(y),\nabla_b\boldsymbol{\Phi}^k(y)]\mathbf{U}^\dagger=0.
\end{equation} 

Indeed, from the family of Lie algebra gauge equivalent domain wall solutions 
$(\tilde{\boldsymbol{\Phi}}^k,\tilde{\boldsymbol{\!A}}^k_a;g^k_{ab})$, we can choose 
a gauge such that 
\[
(\tilde{\boldsymbol{\Phi}}^k,\tilde{\boldsymbol{\!A}}^k_a;g^k_{ab})\mapsto(\boldsymbol{\Phi}^k, \mathbf{0}^{}_a;g^k_{ab}).
\]
However, the search for analytical solutions 
$(\boldsymbol{\Phi}^k, \mathbf{0}^{}_a;g^k_{ab})$ is still a non trivial task. For 
these field configurations  
we will restrict ourselves to consider only those completely integrable models that 
were obtained, for special values of the parameters in the Higgs potential 
(\ref{potential1}), in Ref.\cite{Melfo:2011ev}. These are given by 
 \begin{equation}\label{kink}
\boldsymbol{\Phi}^k(y)=\phi_M(y)\mathbf{M} + \phi_P(y)\mathbf{P},
\end{equation}
\begin{equation}\label{kink_comp}
\phi_M(y)=v\tanh by,\qquad \phi_P(y)=v\kappa,
\end{equation} 
where $\mathbf{M}$ and $\mathbf{P}$ are two conmuting orthogonal diagonal 
generators of $\mathfrak{su}(5)$, and $g^k_{ab}$ given by (\ref{metric-ansatz}) 
with
\begin{equation}\label{warp}
 A(y)=-\frac{v^2}{9}[2 \ln \left(\text{cosh}\,by \right)+\frac{1}{2}\tanh^2 by].
\end{equation}
The space-time is asymptotically $AdS_5$ with cosmological constant 
$\Lambda=-8b^2v^4/27$.  
All the couplings which appear in (\ref{potential1}) can be written explicitly in terms of 
$v$ and $b$. The choice of $\mathbf{M}$ and $\mathbf{P}$ relies  
on the asymptotic values of $\boldsymbol{\Phi}^k$ at 
$y\rightarrow \pm\infty$, which are linked to the possible symmetry breaking 
patterns, and $\kappa$ in (\ref{kink_comp}) is a numerical constant which 
depends on this choice.

As discussed in \cite{Melfo:2011ev} (see \cite{Vachaspati:2001pw,Pogosian:2001fm} 
for the flat space case), by imposing the topologically non-trivial boundary conditions
\begin{eqnarray}\label{boundaryI}
\boldsymbol{\Phi}_{\bf A}^k(+\infty)& \sim &  v\,{\rm diag( \;3, \;3,-2,-2,-2)}, \nonumber \\
\boldsymbol{\Phi}_{\bf A}^k(-\infty) & \sim &  v\,{\rm diag( \;2, \;2,-3,-3,\, \;\;2)},
\end{eqnarray}
a spatially-dependent symmetry breaking pattern is then obtained, where the unbroken symmetries 
$H_{\pm}$ (at $y\rightarrow \pm\infty$) and $H_0$ (at $y=0$) are given by 
\begin{equation}
H^{\bf A}_\pm = \frac{SU(3)_\pm \times SU(2)_\pm \times U(1)_\pm}{Z_3\times Z_2},
\end{equation}
\begin{equation}
 H^{\bf A}_0 = \frac{SU(2)_+ \times SU(2)_- \times U(1)_M \times U(1)_P}{Z_2 \times Z_2},
\end{equation}
with the following embeddings
\begin{equation}
SU(2)_\mp \subset SU(3)_\pm.
\end{equation}
On the other hand, for $\boldsymbol{\Phi}^k$ taking the asymptotic values 
\cite{Melfo:2011ev}
\begin{eqnarray} \label{boundaryII}
\boldsymbol{\Phi}_{\bf B}^k(+\infty)& \sim & v\,{\rm diag( \;\;1, \; \;1, \;\;1, \;\;1,-4)}, \nonumber \\
\boldsymbol{\Phi}_{\bf B}^k(-\infty) & \sim & v\,{\rm diag( -1, -1,-1,4, -1)},
\end{eqnarray}
$SU(5)$ breaks to 
\begin{equation}
 H^{\bf B}_\pm = \frac{SU(4)_\pm \times  U(1)_\pm}{Z_4},
\end{equation}
 \begin{equation}
 H^{\bf B}_0 = \frac{SU(3) \times U(1)_M \times U(1)_P}{Z_3 },
 \end{equation}
where {$SU(3)$} is embedded in different manners in {$SU(4)_+$} and {$SU(4)_-$}.
      
The domain wall configurations 
$(\tilde{\boldsymbol{\Phi}}^k,\tilde{\!\boldsymbol{A}}^k_a;g^k_{ab})$ provide 
regularizations of the Randall-Sundrum brane-world, in which the $SU(5)$ gauge 
symmetry of the theory (\ref{theory1}) is \textit{broken} to a spatially-dependent 
subgroup $H$. On the core of the wall the gauge group $H_0$ is an explicit gauge 
symmetry, while at $y\rightarrow\pm\infty$ (as one approaches the $AdS$ horizons) 
the explicit gauge group is $H_{\pm}$, with $H_0=H_+\cap H_-$ being differently 
embedded in $H_+$ and $H_-$.

It should be noted that, while domain walls in abelian $Z_2$-symmetric theories are 
topologically stable, there is no global stability criterium for the non-abelian ones. This 
lead us to resort to perturbative analyses 
(see \cite{Vachaspati:2001pw,Pogosian:2001fm} for flat 
space and \cite{Pantoja:2015yin} for the gravitating global $SU(5)$ cases) to 
establish at least their perturbative stability. Hence, after 
a domain wall configuration 
$(\tilde{\boldsymbol{\Phi}}^k,\tilde{\!\boldsymbol{A}}^k_a;g^k_{ab})$ is found, its 
perturbative stability, which involves second variations of the action and depends 
explicitly of the field content of the theory, should be addressed.
On the other hand, if the original gauge symmetry is spontaneously 
broken due to this domain wall configuration, we may expect 
that a Higgs mechanism takes place with some imprints on the 4D modes of the 
field fluctuations. 
 
\section{Fluctuations of the Domain Wall configuration}\label{fluctuations}
\subsection{Diffeomorphisms, Lie algebra gauge transformations and fluctuations}

For the determination of the stability of the domain wall solutions and the analysis 
of the gravitationally trapped content on their cores, we shall consider perturbative 
expansions to first order in the fluctuations around the domain wall background.

Let us briefly review the procedure chosen to obtain the perturbation equations, 
which is applicable to any covariant field theory. Consider the set
\begin{equation}\label{proc1}
\mathcal{E}[\boldsymbol{\Phi},\boldsymbol{A}_a;g_{ab}]=0
\end{equation}
of field equations (\ref{field_eq1},\ref{field_eq2},\ref{field_eq3},\ref{field_eq4}) of the 
theory (\ref{theory1}) and let 
$(^0\boldsymbol{\Phi},^0\!\!\boldsymbol{A}_a; ^0\!\!g_{ab})$ be a solution of the set 
$\mathcal{E}$. Now, suppose there exists a one-parameter family of solutions 
$(\boldsymbol{\Phi}{(\lambda)},\boldsymbol{A}_a{(\lambda)};g_{ab}{(\lambda)})$, 
\begin{equation}\label{proc2}
\mathcal{E}[\boldsymbol{\Phi}{(\lambda)},\boldsymbol{A}{(\lambda)}_a;g_{ab}{(\lambda)}]=0,
\end{equation}
such that $(\boldsymbol{\Phi}{(0)},\boldsymbol{A}{(0)}_a;g_{ab}{(0)})= (^0\boldsymbol{\Phi},^0\!\!\boldsymbol{A}_a;^0\!\!g_{ab})$. Provided that suitable 
differentiability conditions for $\mathcal{E}$ and 
$(\boldsymbol{\Phi}{(\lambda)},\boldsymbol{A}{(\lambda)}_a;g_{ab}{(\lambda)})$ 
hold, we have
\begin{equation}\label{proc3}
\left.\frac{d}{d\lambda}\mathcal{E}[\boldsymbol{\Phi}{(\lambda)},\boldsymbol{A}{(\lambda)}_a;g_{ab}{(\lambda)}]\right|_{\lambda=0}= 0,
\end{equation}
comprising a set of linear equations for
\begin{equation}\label{proc4}
\boldsymbol{\varphi}= \left.\frac{d}{d\lambda}\boldsymbol{\Phi}{(\lambda)}\right|_{\lambda=0},\quad \boldsymbol{\mathcal{A}}_a=\left.\frac{d}{d\lambda}\boldsymbol{A}_a{(\lambda)}\right|_{\lambda=0}
\end{equation}
and
\begin{equation}\label{proc5}
h_{ab}=\left.\frac{d}{d\lambda}g_{ab}{(\lambda)}\right|_{\lambda=0},
\end{equation}
which are the scalar, vector gauge and metric fluctuations, respectively, around 
the background given by $(^0\boldsymbol{\Phi},^0\!\!\boldsymbol{A}_a;^0\!\!g_{ab})$.

Now, from (\ref{proc4},\ref{proc5}), it follows that under an infinitesimal 
diffeomorphism
\begin{equation}\label{diffeomorphism}
 x^{a} \mapsto x^{a}+\epsilon^{a},
 \end{equation}
we have
\begin{equation}\label{change2}
\boldsymbol{\varphi}\mapsto \boldsymbol{\varphi}+\pounds_{\epsilon}{^0\boldsymbol{\Phi}},\qquad
\boldsymbol{\mathcal{A}}_a\mapsto \boldsymbol{\mathcal{A}}_a +\pounds_{\epsilon}{^0\!\boldsymbol{A}}_a
\end{equation}
and
\begin{equation}\label{change1}
h_{ab}\mapsto h_{ab}+\pounds_{\epsilon}{^0\!g_{ab}},
\end{equation}
where $\pounds_{\epsilon}$ is the Lie derivative with respect to the vector field 
$\epsilon^a$. The full space-time diffeomorphism invariance of the theory 
(\ref{theory1}) implies that 
$(\boldsymbol{\varphi},\boldsymbol{\mathcal{A}}_a,h_{ab})$ and 
$({\boldsymbol{\varphi}}+\pounds_{\epsilon}{^0\boldsymbol{\Phi}},{\boldsymbol{\mathcal{A}}}_a+\pounds_{\epsilon}{^0\!\boldsymbol{A}}_a,{h}_{ab}+\pounds_{\epsilon}{^0\!g_{ab})}$ 
describe the same physical perturbations. 

On the other hand, (\ref{theory1}) is also invariant under Lie algebra gauge transformations. It follows from (\ref{proc4},\ref{proc5}) that under infinitesimal 
Lie algebra gauge transformations we have
\begin{equation}\label{change4}
\boldsymbol{\varphi}\mapsto \boldsymbol{\varphi} -i [\boldsymbol{\sigma},{^0\boldsymbol{\Phi}}],\qquad
{\boldsymbol{\mathcal{A}}}_a\mapsto {\boldsymbol{\mathcal{A}}}_a + \frac{1}{\texttt{g}}{^0\mathbf{D}}_a\boldsymbol{\sigma},\end{equation}
and
\begin{equation}\label{change5}
 h_{ab}\mapsto h_{ab},\quad
\end{equation}
where $\boldsymbol{\sigma}$ is a Lie algebra valued scalar field parameterizing 
the gauge freedom and $^0\mathbf{D}_a$ is the gauge covariant derivative with respect the background gauge field $^0\!\boldsymbol{A}_a$.

\subsection{(4+1) decomposition of the fluctuations}

For a background 
$({^0\boldsymbol{\Phi}},{^0\!\boldsymbol{A}}_a;{^0\!g}_{ab})$ that preserves 
4D-Poincare invariance, it is convenient decompose $h_{ab}$ as \cite{Giovannini:2001fh}
\begin{eqnarray}\label{5_tensor}
h_{ab}=&&2e^{2A}\left(h_{\mu\nu}^{TT}+\partial_{(\mu}f_{\nu)}+\eta_{\mu\nu}\psi+\partial_{\mu}\partial_{\nu}E\right)dx^\mu_adx^\nu_b \nonumber\\
&&+ e^A\left(D_{\mu}+\partial_{\mu}C\right)\left(dx^\mu_ady^{}_b + dy^{}_adx^\mu_b\right)\nonumber\\&&+ 2\omega \,dy^{}_ady^{}_b,
\end{eqnarray}
where
\begin{equation}
h_{\mu}^{TT}{}^{\mu}=0,\quad
\partial^{\mu}h_{\mu\nu}^{TT}=0
\end{equation}
and
\begin{equation}
\partial^{\mu}f_{\mu}=0,\quad
\partial^{\mu}D_{\mu}=0.
\end{equation}
We may also set 
\begin{equation}\label{gauge_field_decomposition}
\boldsymbol{\mathcal{A}}_a=\boldsymbol{\mathcal{A}}_\mu dx^\mu_a +\boldsymbol{\mathcal{A}}_y dy^{}_a.
\end{equation}

Now, for an infinitesimal diffeomorphism (\ref{diffeomorphism}) of the form
\begin{equation}\label{diffeo1} 
\epsilon_a=e^{2A}\epsilon_{\mu}dx^\mu_a +\epsilon_y dy_a,
\end{equation}
where
\begin{equation}\label{diffeo2}
\epsilon_{\mu}=\partial_{\mu}\epsilon+\zeta_{\mu},\qquad
\partial^{\mu}\zeta_{\mu}=0,
\end{equation}
we have that (\ref{change1}) induce the following transformations
\begin{equation}
\psi\mapsto\psi-A'\epsilon_{y},\qquad\omega\mapsto \omega+\partial_y\epsilon_{y},
\end{equation}
\begin{equation}
E\mapsto E-\epsilon,\qquad
C\mapsto C-e^{A}\partial_y\epsilon+e^{-A}\epsilon_{y},
\end{equation}
\begin{equation}
D_\mu\mapsto D_{\mu}-e^{A}\partial_y\zeta_{\mu},\qquad f_\mu\mapsto f_{\mu}-\zeta_{\mu}, 
\end{equation}
and
\begin{equation}\label{hTT}
{h}_{\mu\nu}^{TT}\mapsto h_{\mu\nu}^{TT},
\end{equation}
where a prime $'$ denotes the derivative with respect to $y$. Indeed, at this point all these fields depend not only on the point $x^\mu$ in the four space but also on the coordinate $y$ along the additional dimension.

As follows from (\ref{hTT}), $h_{\mu\nu}^{TT}$ is automatically diffeomorphism 
invariant. The next step is to
complete an appropriate set of quantities which are invariant under infinitesimal 
diffeomorphisms. One may use the above transformations to construct the 
vector field $u^a$ given by
\begin{equation}
u^a\equiv (\partial^\mu E + f^\mu){\partial}_\mu^a + (e^{2A}E'-e^A C)\partial_y^a,
\end{equation}
which, under an infinitesimal diffeomorphism (\ref{diffeomorphism}) of the form (\ref{diffeo1},\ref{diffeo2}), 
transforms as
\begin{equation}
u^a\mapsto u^a -\epsilon^a.
\end{equation}
Hence, since $\pounds_u$ is linear with respect to $u^a$, the quantities
\begin{equation}\label{invariant_metric_fluctuation} 
h^{\text{inv}}_{ab}\equiv h_{ab}+\pounds_{u}{^0\!g}_{ab},
\end{equation}
\begin{equation}\label{invariant_scalar_fluctuation}
\boldsymbol{\varphi}^{\text{inv}}\equiv\boldsymbol{\varphi}+\pounds_{u}{^0\boldsymbol{\Phi}}
\end{equation}
and
\begin{equation}\label{invariant_gauge_fluctuation}
\boldsymbol{\mathcal{A}}_a^{\text{inv}}\equiv\boldsymbol{\mathcal{A}}_a+\pounds_{u}{^0\!\boldsymbol{A}_a},
\end{equation}
are invariant under an infinitesimal diffeomorphism 
(\ref{diffeomorphism},\ref{diffeo1},\ref{diffeo2}).
 
In particular, from (\ref{invariant_metric_fluctuation}) we find that 
\begin{eqnarray}\label{5_tensor_inv}
h^{\text{inv}}_{ab}=&&2e^{2A}\left(h_{\mu\nu}^{TT}+\eta_{\mu\nu}\psi^{\text{inv}}\right)dx^\mu_adx^\nu_b \nonumber\\
&&+ e^A D^{\text{inv}}_{\mu}\left(dx^\mu_ady^{}_b + dy^{}_adx^\mu_b\right)\nonumber\\&&+ 2\,\omega^{\text{inv}} dy^{}_ady^{}_b,
\end{eqnarray}
where
\begin{equation}
{\psi}^{\text{inv}}\equiv{\psi}-A'\left(e^{2A}{E}'-e^{A}{C}\right),
\end{equation} 
\begin{equation}\label{invariant_vector}
{D}_{\mu}^{\text{inv}}\equiv{D}_{\mu}-e^{A}{f}'_{\mu}.
\end{equation}
and
\begin{equation}
{\omega}^{\text{inv}}\equiv{\omega} + \left(e^{2A}{E}'-e^{A}{C}\right)'.
\end{equation}

Notice that in the generalized longitudinal gauge, $E=C=0$ and $f_\mu=0$, the 
freedom of the coordinate transformations 
(\ref{diffeomorphism},\ref{diffeo1},\ref{diffeo2}) is completely fixed and the 
diffeomorphism-invariant fluctuations coincide with the original ones, i.e.
\[
h_{ab}^{\text{inv}}=h_{ab},\qquad \boldsymbol{\varphi}^{\text{inv}}=\boldsymbol{\varphi}\qquad \boldsymbol{\mathcal{A}}_a^{\text{inv}}=\boldsymbol{\mathcal{A}}_a.
\] 
Thus, in the generalized 
longitudinal gauge, the evolution equations satisfied by the field fluctuations $h_{ab}$, 
${\boldsymbol{\varphi}}$ and ${\boldsymbol{\mathcal{A}}_a}$ also hold  for 
the diffeomorphism-invariant fluctuations $h_{ab}^\text{{inv}}$, 
${\boldsymbol{\varphi}}^\text{{inv}}$ and ${\boldsymbol{\mathcal{A}}_a}^\text{{inv}}$. 
Since only diffeomorphism-invariant fluctuations will be considered, we shall in the 
following drop the superscript $\text{inv}$ on these ones.

On the other hand, as follows from the Lie algebra gauge invariance of the theory 
(\ref{change4},\ref{change5}), the field fluctuations 
$(\boldsymbol{\varphi}, \boldsymbol{\mathcal{A}}_a,h_{ab})$ and 
$(\boldsymbol{\varphi}-i [\boldsymbol{\sigma},{\boldsymbol{\Phi}}^k], \boldsymbol{\mathcal{A}}_a + {\texttt{g}}^{-1}\nabla_a\boldsymbol{\sigma},h_{ab})$, with 
$\boldsymbol{\sigma}$ a Lie algebra valued scalar field parameterizing the gauge 
freedom, describe the same physical perturbations.When gauge field localization 
on domain walls is discussed, it is often considered a gauge fixing in which the extra 
dimension component $\boldsymbol{\mathcal{A}}_y$ of the gauge field 
$\boldsymbol{\mathcal{A}}_a$ vanishes. Here, we will consider instead 
field fluctuations which do not change under Lie algebra gauge transformations, since in terms of these the obtained results will be independent of any gauge fixing.  
In the following, the dependence on the additional coordinate will be expressed in the conformal coordinate $z$
\begin{equation}\label{y_zeta}
dy_a= e^{A(z)}dz_a,
\end{equation}
such that
\begin{equation}
g^k_{ab}=e^{2A(z)}\left(\eta_{\mu\nu}dx^{\mu}_adx^{\nu}_b +dz_a dz_b\right).
\end{equation}

Let $\boldsymbol{\mathcal{A}}_a$ be the gauge vector 
fluctuation (\ref{gauge_field_decomposition}) with
\begin{equation}\label{gauge_T_L}
\boldsymbol{\mathcal{A}}_\mu=e^{-A/2}\boldsymbol{a}_\mu + \boldsymbol{\mathcal{A}}_\mu^L,
\end{equation}
where
\begin{equation}
\partial^\mu\boldsymbol{a}_\mu=0,\qquad \boldsymbol{\mathcal{A}}_\mu^L=\partial_\mu\boldsymbol{\chi}.
\end{equation}
The Lie algebra gauge- and diffeomorphism-invariant fluctuations we use are 
$\boldsymbol{\alpha}$, $\boldsymbol{\beta}$ and $\boldsymbol{a}_\mu$, where
\begin{equation}
\boldsymbol{\alpha}\equiv \boldsymbol{\varphi}+i\texttt{g}[\boldsymbol{\chi},\boldsymbol{\Phi}^k],
\end{equation}
and
\begin{equation}\boldsymbol{\beta}\equiv e^{-A(z)/2}(\boldsymbol{\mathcal{A}}_z-\partial_z\boldsymbol{\chi}),
\end{equation}
together with $h_{\mu\nu}^{TT}$, $D_\mu$, $\psi$ and $\omega$ which are unchanged under a Lie algebra gauge transformation (\ref{change5}).

\subsection{Linearized perturbation equations} 

Making the $(4+1)$ decomposition discussed in the previous subsection, from the 
set of linearized field equations for the fluctuations $\boldsymbol{\varphi}$, 
$\boldsymbol{\mathcal{A}}_a$ and $h_{ab}$ around the domain wall background 
(see Appendix \ref{appendixA}), we obtain the field equations for the chosen set of 
diffeomorphism- and Lie algebra gauge-invariant fluctuations. We find 
(where now and in the following a prime $'$ denotes the derivative with respect to 
$z$)
\begin{equation}\label{tensor_TT}
\left(\partial^\rho\partial_\rho+3A'\partial_z +\partial_z^2\right) h_{\mu\nu}^{TT}=0,
\end{equation}
\begin{equation}\label{vector}
\partial_{(\mu}(\partial_z+3A')D_{\nu)}=0,\quad \partial^\nu\partial_\nu D_\mu=0,
\end{equation}
\begin{eqnarray}
-(\partial^\rho\partial_\rho\psi+\partial_z^2\psi+7A'\partial_z\psi)+(6{A'}^2+2A'')\omega\nonumber\\+A'\partial_z\omega=e^{2A}\left.\frac{2}{3}\frac{\partial V(\boldsymbol{\Phi})}{\partial\phi_q}\right|_{\boldsymbol{\Phi}^k}\varphi_q,\qquad
\end{eqnarray}
\begin{eqnarray}
-\left(\partial^\mu\partial_\mu\omega+(6A'^2+2A'')\omega+4A'\partial_z\omega\right)-2\phi_M'\partial_z\varphi_M\nonumber\\-4\left(A'\partial_z\psi+\partial_z^2\psi\right)=e^{2A}\left.\frac{2}{3}\frac{\partial V(\boldsymbol{\Phi})}{\partial \phi_q}\right|_{\boldsymbol{\Phi}^k}\varphi_q,\quad
\end{eqnarray}
and the two constraints
\begin{equation}
\partial_\mu(\omega+2\psi)=0, \quad \partial_\mu\left(3A'\omega-3\partial_z\psi-\phi_M'\varphi_M^{}\right)=0.
\end{equation}
Also we find
\begin{eqnarray}\label{scalar_fluctuations_y}
e^{-2A}&&\partial^\mu\partial_\mu\boldsymbol{\alpha} +e^{-5A}\partial_z\left(e^{3A}\partial_z{\boldsymbol{\alpha}}\right) - \left.\frac{\partial^2 V(\boldsymbol{\Phi})}{\partial\phi_p\partial\phi_q}\right|_{\boldsymbol{\Phi}^k}\alpha_p\mathbf{T}^q\nonumber\\ &&+ 4e^{-2A}\partial_z{\boldsymbol{\Phi}^k}\partial_z\psi - 2e^{-5A}\partial_z\left(e^{3A}\partial_z\boldsymbol{\Phi}^k\right)\omega\nonumber\\&& -e^{-2A}\partial_z{\boldsymbol{\Phi}^k}\partial_z\omega=-2i\texttt{g}e^{-2A}[\boldsymbol{\beta},\partial_z{\boldsymbol{\Phi}^k}]\nonumber\\&&\quad\qquad-i\texttt{g}e^{-5A}[\partial_z(e^{3A}\boldsymbol{\beta}),\boldsymbol{\Phi}^k],
\end{eqnarray}
\begin{equation}\label{eom_beta}
e^{-2A}\partial^\mu\partial_\mu\boldsymbol{\beta}- (M^2)^{qp}\beta_p\mathbf{T}^q=i\texttt{g}[\boldsymbol{\Phi}^k,\partial_z\boldsymbol{\alpha}] + i\texttt{g}[\boldsymbol{\alpha},\partial_z\boldsymbol{\Phi}^k],
\end{equation}
with the constraint
\begin{equation}\label{alpha_beta_constraint}
e^{-3A}\partial_z\left(e^A\boldsymbol{\beta}\right)=-i\texttt{g}[\boldsymbol{\Phi}^k,\boldsymbol{\alpha}],
\end{equation}
and
\begin{eqnarray}\label{gauge_mu}
\!\!\!\!\!\!e^{-2A}\left(\partial^\mu\partial_\mu\boldsymbol{a}_\nu-\left(\frac{1}{4}A'^2 + \frac{1}{2}A''\right)\boldsymbol{a}_\nu +\partial_z^2\boldsymbol{a}_\nu\right)\nonumber\\ -(M^2)^{qp}(\boldsymbol{a}_\nu)_p\mathbf{T}^q=0,
\end{eqnarray}
where the matrix ${M}^2$ is given by
\begin{equation}\label{y_dependent_mass}
({M}^2)^{qp}=-2\texttt{g}^2\text{Tr}\left\{[\mathbf{T}^q,\boldsymbol{\Phi}^k(y)][\mathbf{T}^p,\boldsymbol{\Phi}^k(y)]\right\}.
\end{equation}

As follows from (\ref{tensor_TT}), (\ref{vector}) and (\ref{gauge_mu}), the fluctuations  
$h_{\mu\nu}^{TT}$, $D_\mu$ and $\boldsymbol{a}_\mu$ decouple one of each other 
as also from the rest of the field fluctuations, with  
$h_{\mu\nu}^{TT}$ and $D_\mu$ behaving as its corresponding analogous in 
global $SU(5)\times Z_2$ \cite{Pantoja:2015yin} and the standard abelian $Z_2$ 
\cite{Giovannini:2001fh} domain walls. It follows that the tensor perturbation 
$h_{\mu\nu}^{TT}$ 
has no tachyonic modes which destabilize the domain wall background, there is a 
normalizable massless mode which gives rise to 4D gravity on the core of the wall 
and a tower of non-normalizable massive KK modes which propagate in the bulk. 
There is no localized vector fluctuation $D_\mu$. We refer the reader to 
Ref.\cite{Pantoja:2015yin} and references therein for a detailed discussion. In the 
following, we will restrict ourselves to the analysis of the scalar field fluctuations 
$\psi$, $\omega$, $\boldsymbol{\alpha}$ and $\boldsymbol{\beta}$ and the vector 
field fluctuation $\boldsymbol{a}_\mu$. 

\subsection{Lie algebra decomposition of the fluctuations}

Let us consider the following Cartan decomposition of the $\mathfrak{su}(5)$ 
Lie algebra 
\begin{equation}
\mathfrak{su}(5)=\mathcal{K}\oplus\mathcal{K}^\bot,
\end{equation}
where the Lie subalgebra $\mathcal{K}$ is given by
\begin{equation}
\mathcal{K}=\{\mathbf{T}_0\}\oplus\{\mathbf{T}_{\text{br}}\}
\end{equation}
with the subsets $\{\mathbf{T}_0\}$ and $\{\mathbf{T}_{\text{br}}\}$ 
defined by
\begin{equation}\label{To}
[\mathbf{T}_0^q,\boldsymbol{\Phi}^k(y)]=0, \quad q=1,\ldots,n_0,
\end{equation}
and
\begin{equation}
[\mathbf{T}_{\text{br}}^q,\boldsymbol{\Phi}^k(y)]\neq 0, \quad q=1,\ldots,n_{\text{br}}. 
\end{equation} 
The orthogonal complement $\mathcal{K}^\bot$ is given by
\begin{equation}
\mathcal{K}^\bot=\{\mathbf{T}_+\}\oplus\{\mathbf{T}_-\},
\end{equation}
where  
\begin{equation}\label{T_plus}
[\mathbf{T}_+^q,\boldsymbol{\Phi}^k(\infty)]=0,\,\, [\mathbf{T}_+^q,\boldsymbol{\Phi}^k(-\infty)]\neq 0,\quad q=1,\ldots,n_+,
\end{equation}
and 
\begin{equation}\label{T_minus}
[\mathbf{T}_-^q,\boldsymbol{\Phi}^k(-\infty)]=0,\,\, [\mathbf{T}_-^q,\boldsymbol{\Phi}^k(\infty)]\neq 0,\quad q=1,\ldots,n_-.
\end{equation}
We have $n_0=8$, $n_{\text{br}}=8$ and $n_+=n_-=4$ for the symmetry breaking 
$\bf{A}$. On the other hand, $n_0=10$, $n_{\text{br}}=2$ and $n_+=n_-=6$ for the symmetry breaking 
$\bf{B}$. 
 
Let $\left.\partial^2 V(\boldsymbol{\Phi})/\partial\phi_q^2\right|_{\boldsymbol{\Phi}^k}^{\{i\}}$ be the restriction 
of the hessian of $V(\boldsymbol{\Phi})$ at $\boldsymbol{\Phi}^k$ to the subspace spanned
by the subset $\{\mathbf{T}_i\}$. First, let us consider 
$\left.\partial^2 V(\boldsymbol{\Phi})/\partial\phi_q^2\right|_{\boldsymbol{\Phi}^k}^{\{0\}}$. For $q=M$, we find
\begin{eqnarray}\label{potential_MM}
\left.\frac{\partial^2 V(\boldsymbol{\Phi})}{{\partial\phi_M}^2}\right|_{\boldsymbol{\Phi}^k}^{\{0\}}=&&-2b^2(1+\frac{2}{3}v^2)\nonumber\\&&+b^2F^2\left(6+\frac{4}{3}v^2(4-\frac{5}{3}F^2)\right),
\end{eqnarray}
where 
\begin{equation}\label{F}
F=\tanh by,\quad y=y(z),
\end{equation}
and for $q=P$ we have
\begin{equation}\label{potential_PP}
\left.\frac{\partial^2 V(\boldsymbol{\Phi})}{{\partial\phi_P}^2}\right|_{\boldsymbol{\Phi}^k}^{\{0\}}= 4b^2(1+ \frac{4}{9}v^2)\equiv \mathcal{M}^2_{H},
\end{equation}
in both symmetry breaking patterns.  

For $q$ such that $\mathbf{T}_0^q\neq \mathbf{M}, \mathbf{P}$, we find for the symmetry breaking {\bf A} 
\begin{eqnarray}\label{V_H_0_A}
\!\!\!\!\!\left.\frac{\partial^2 V(\boldsymbol{\Phi})}{{\partial\phi_q}^2}\right|_{\boldsymbol{\Phi}^k}^{\{0\}}=&&b^2\left(10 + \frac{20}{9}v^2\right)\nonumber\\&& -b^2F\left(\frac{4}{3}v^2F\mp(6-\frac{4}{9}v^2(F^2+1))\right)
\end{eqnarray}
where the $\pm$ signs correspond to the two different 
$SU(2)$ in $H_{0}^{\boldsymbol{A}}$. In 
the symmetry breaking {\bf B} we find 
\begin{eqnarray}\label{V_H_0_B}
\left.\frac{\partial^2 V(\boldsymbol{\Phi})}{{\partial\phi_q}^2}\right|^{\{0\}}_{\boldsymbol{\Phi}^k}=&& b^2\left( \frac{5}{2} + \frac{5}{4}v^2\right)\nonumber\\&&+ b^2F^2\left(\frac{3}{2} + \frac{1}{6}v^2(5 -\frac{11}{6} F^2)\right).
\end{eqnarray}

Next, along the subset $\{\mathbf{T}_{\text{br}}\}$, we find 
\begin{equation}\label{potential_br}
\left.\frac{\partial^2 V(\boldsymbol{\Phi})}{{\partial\phi_q}^2}\right|_{\boldsymbol{\Phi}^k}^{\{\text{br}\}}=0,
\end{equation}
in both symmetry breaking patterns. Along the subsets $\{\mathbf{T}_+\}$ and 
$\{\mathbf{T}_-\}$ we have 
\begin{equation}\label{TpmA}
\left.\frac{\partial^2 V(\boldsymbol{\Phi})}{{\partial\phi_q}^2}\right|_{\boldsymbol{\Phi}^k}^{\{\pm\}}= 2b^2F\left(1 +\frac{2}{3}v^2(1- \frac{1}{3} F^2)\right)(F\pm 1)
\end{equation}
for both symmetry breaking patterns. The field fluctuations 
$\boldsymbol{\varphi}\in \{\mathbf{T}_{\text{br}}\}$ are 5D 
Nambu-Goldstone bosons, while the remaining scalar field fluctuations correspond 
to the 5D massive scalar fields (in general, with $y$-dependent \textit{masses}) of the 
spontaneously broken gauge theory.   

Now, let $({M}^2)^{\{i\}}$ be the restriction of $M^2$, as given by 
(\ref{y_dependent_mass}), to the subspace spanned by the subset 
$\{\mathbf{T}_i\}$. We have
\begin{equation}
({M}^2)^{\{i\}}= \mathcal{M}_{i}^2\mathbb{I}^{\{i\}},
\end{equation}
where $\mathbb{I}^{\{i\}}$ is the identity matrix of dimension $n_i\times n_i$,  
\begin{equation}\label{mass_w_0}
\mathcal{M}_{0}^2=0,
\end{equation} 
\begin{equation}\label{mass_w_1}
\mathcal{M}^2_{{\rm br}}= \mathcal{M}^2_{W},\quad \mathcal{M}^2_{W}\equiv\frac{5}{2}v^2\texttt{g}^2, 
\end{equation}
and
\begin{equation}\label{mass_w_2}
\mathcal{M}^2_{\pm}= \frac{1}{4}\mathcal{M}_W^2(1 \mp F)^2,
\end{equation}
for the symmetry breaking $\mathbf{A}$,
with essentially the same results for the symmetry breaking 
$\mathbf{B}$ differing only in numerical factors. $\mathcal{M}^2_i$ are the gauge boson 5D \textit{masses} generated through the Higgs mechanism.

\section{Dimensional reduction}\label{4D_modes}

\subsection{Fluctuations along $\{\mathbf{T}_0\}$}

The gauge invariant fluctuations  
$(\boldsymbol{\alpha}, \boldsymbol{\beta},\boldsymbol{a}_\mu)\in \{\mathbf{T}_0\}$ 
decouple one of each other and $\boldsymbol{\alpha}=\boldsymbol{\varphi}$, i.e. the 
\textit{wall fluctuations} $\boldsymbol{\varphi}$ are gauge invariant in this sector. Let 
us recall that the theory considered maintains an explicit $H_0$ gauge symmetry. On 
the other hand, the \textit{gravitational} fluctuations $\psi$ and $\omega$ mixes with 
$\boldsymbol{\alpha}$. We find the constraints
\begin{equation}\label{constraint1}
2\psi + \omega=0\qquad
3A'\omega -3\partial_z\psi - \phi_M'\alpha_M=0.   
\end{equation} 
Hence, $\psi$, $\omega$ and $\alpha_M$  are not 
independent and correspond to a single physical scalar fluctuation. Note that 
$\alpha_M$ in (\ref{constraint1}) is associated to a $U(1)$ factor of the subgroup 
$H_0$.

Let $\boldsymbol{\Xi}$ be the scalar fluctuation defined as
\begin{equation}\label{xi}
\boldsymbol{\Xi}\equiv e^{3A/2}\left(\boldsymbol{\alpha}-\frac{\psi}{A'}(\boldsymbol{\Phi}^k)'\right).
\end{equation}
From (\ref{scalar_fluctuations_y}), the Kaluza-Klein (KK) modes 
$\Xi_q(x,z)\sim e^{ip\cdot x}\Xi_q(z)$, with 
$\boldsymbol{\Xi}\in  \{\mathbf{T}_0\}$, satisfy Schr\"odinger-like equations which 
depends on $q$
\begin{equation}\label{Xi_non_M}
(-\partial_{z}^2 + V_{q}^{\{0\}}) \,\Xi_q= m^2\Xi_q,
\end{equation}
where $V_q^{\{0\}}$ is given by
\begin{equation}\label{Vq}
V_{q}^{\{0\}}= V_{Q_1} + 
\left.e^{2A}\frac{\partial^2V(\boldsymbol{\Phi})}{\partial\phi_q^2}\right|_{\boldsymbol{\Phi}^k}^{\{0\}},
\end{equation}
with $V_{Q_1}$ given by
\begin{equation}\label{VQ1}
V_{Q_1}=\frac{9}{4}A'^{2}+\frac{3}{2}A'' 
\end{equation}
and $p_\mu p^\mu=-m^2$.

Now, the Schr\"odinger operator in (\ref{Xi_non_M}) can be rewritten as
\begin{eqnarray}\label{schro_scalar}
-\partial_{z}^2 + V_q^{\{0\}}=&&\left(\partial_z+\frac{3}{2}A'\right)\left(-\partial_z+\frac{3}{2}A'\right)\nonumber \\&& +\left.e^{2A}\frac{\partial^2V(\boldsymbol{\Phi})}{\partial\phi_q^2}\right|_{\boldsymbol{\Phi}^k}^{\{0\}},
\end{eqnarray}
where the first term of the right hand side is a nonnegative definite operator on 
normalizable functions. If the second term is never negative, the eigenvalues of 
(\ref{Xi_non_M}) are always nonnegative and there are no tachyonic modes.

For $q=M$, $\left.\partial^2 V/\partial\phi^2_M\right|^{\{0\}}_{\boldsymbol{\Phi}^k}$ is 
given by (\ref{potential_MM}) which is no everywhere positive. However, 
$V_M^{\{0\}}$ can be 
written entirely in terms of the derivatives of $\phi_M$ and $A$ using their equations 
of motion \cite{Pantoja:2015yin}. In this case, the Schr\"odinger operator can be 
rewritten as
\begin{equation}\label{super2}
(-\partial_{z}^{2}+V_{M}^{\{0\}})=\left(\partial_{z}+(\ln Z)'\right)
\left(-\partial_{z} + (\ln Z)'\right),
\end{equation}
where
\begin{equation}
Z(z)=e^{3A/2}\frac{\phi'_M}{A'}.
\end{equation}
The corresponding Schr\"odinger-like equation admits no mode with $m^2<0$, a 
massless solution $\Xi_M^0(z)\propto Z(z)$, 
which is no normalizable since $Z(z)$ is not bounded for $z\rightarrow 0$, and  
a tower of continuum states with $m^2>0$. Since $V_M^{\{0\}}$ has the shape 
of a symmetric potential barrier of infinite height, the continuum of massive modes 
behave as waves that propagates in the bulk being repelled off the core of the wall. 
The behavior of the scalar fluctuation $\Xi_M$ close parallels the one of the scalar 
fluctuation associated to the standard abelian $Z_2$ kink \cite{Giovannini:2001fh}.

Next, let us consider the scalar perturbations $\Xi_q$ along the generators 
$\mathbf{T}_0^q$ other than $\mathbf{M}$.
For the scalar fluctuation $\Xi_P$ along the generator $\mathbf{P}$, associated to 
the $U(1)_P$ in $H_0$, 
$\left.\partial^2 V/\partial\phi^2_P\right|^{\{0\}}_{\boldsymbol{\Phi}^k}$ is a positive 
definite constant $\mathcal{M}_H^2$ given by 
(\ref{potential_PP}) and $V_P^{\{0\}}$ is everywhere positive with the shape of a 
symmetric potential barrier of finite height. Hence, $V_P^{\{0\}}$ does not support 
bound states with $m^2\leq 0$, while those modes with $m^2>0$ behave as  
scattered waves by the wall. 

For $q\neq M,P$, 
$\left.\partial^2 V/\partial\phi^2_q\right|_{\boldsymbol{\Phi}^k}^{\{0\}}$ is given by 
(\ref{V_H_0_A}) for the symmetry breaking {\bf A} and (\ref{V_H_0_B}) for the 
symmetry breaking {\bf B}. In both symmetry breaking patterns, $\left.\partial^2 V/\partial\phi^2_q\right|_{\boldsymbol{\Phi}^k}^{\{0\}}$ is positive definite and $V_{q}^{\{0\}}$ has the shape of a symmetric potential barrier of finite height, therefore (\ref{schro_scalar}) does not support bound states with $m^2\leq 0$. 

For $\boldsymbol{\beta}$, from (\ref{eom_beta}) and (\ref{alpha_beta_constraint}) 
we find 
\begin{equation}
e^{3A}\partial_z(e^{3A/2}\boldsymbol{\beta})=0=e^{-3A/2}\partial^\mu\partial_\mu\boldsymbol{\beta},
\end{equation}
i.e. $\boldsymbol{\beta}=0$. Therefore, $\boldsymbol{\mathcal{A}}_z=\partial_z\boldsymbol{\chi}$.

Finally, let us consider the gauge vector fluctuations $\boldsymbol{a}_\mu$. From 
(\ref{gauge_mu}) it follows that the modes 
$\boldsymbol{a}_\mu(x,z)\sim e^{ip\cdot x}\boldsymbol{a}_\mu(z)$, with 
$\boldsymbol{a}_\mu\in \{\mathbf{T}_0\}$, satisfy 
the Schr\"odinger-like equation
\begin{equation}\label{gauge_QM_1}
(-\partial_{z}^2 + \mathcal{V}_1)\boldsymbol{a}_{\mu}= m^2\boldsymbol{a}_{\mu},
\end{equation}
where
\begin{equation}\label{VQ2}
\mathcal{V}_{1}=\frac{1}{4}A'^{2}+\frac{1}{2}A'' 
\end{equation}
and $p_\mu p^\mu=-m^2$. In this case, the Schr\"odinger operator can be 
factorized as
\[
\left(-\partial^2_z + \frac{1}{4}A'^2+\frac{1}{2}A''\right)=\left(\partial_z + \frac{1}{2}A'\right) \left(-\partial_z + \frac{1}{2}A'\right).
\]
Hence, (\ref{gauge_QM_1}) admits no modes with $m^2<0$, a massless mode 
\begin{equation}\label{gauge_boson_zero}
(\boldsymbol{a}_\mu)^0_q\sim e^{A/2}\varepsilon_\mu,\qquad p^\mu\varepsilon_\mu=0,
\end{equation}
and a tower of massive modes which propagates in the bulk. 

Note that the massless mode correspond to a constant 
$\boldsymbol{\mathcal{A}}_\mu$ along the additional coordinate. But, as is well 
known from the $Z_2$-symmetric standard kink, a constant zero mode of a vector 
field does not gives rise to a localizable 4D massless vector field \cite{Bajc:1999mh}. 
In order to examine the massive modes, let us suppose a higher enough $AdS_5$ 
curvature $\Lambda$ such that we can approximate $A(z)$ by its thin wall 
or \textit{brane} limit\footnote{The thin wall limit can be defined as the limit $b\rightarrow\infty$ and 
$v\rightarrow 0$, while $|\Lambda|= 8b^2v^4/27$ is kept finite.}
\begin{equation}\label{thin_wall_limit}
A(z)\sim -\ln(1+k|z|),\quad k\equiv (2v^2/9)|b|=\sqrt{-\Lambda/6}. 
\end{equation}
For $m^2\neq0$, in this approximation, the Neumann boundary conditions at $z=\pm\infty$ gives for the even modes
\begin{equation}\label{massive_br_fluct}
(\boldsymbol{a}_\mu)_q\sim \xi^{1/2}\left[Y_1(m\xi) -\frac{Y_0(m/k)}{J_0(m/k)}J_1(m\xi) \right]\varepsilon_\mu,
\end{equation}
where 
\begin{equation}\label{bessel_arg}
\xi=\xi(z)= k^{-1} +|z|,
\end{equation}
with $J_\nu$ and $Y_\nu$ Bessel functions of order $\nu$. The odd modes are 
given by 
\begin{equation}
(\boldsymbol{a}_\mu)_q\sim \xi^{1/2}\left[Y_1(m\xi) -\frac{Y_1(m/k)}{J_1(m/k)}J_1(m\xi) \right]\varepsilon_\mu,\quad z>0,
\end{equation}
with $\boldsymbol{a}_\mu(z)= -\boldsymbol{a}_\mu(-z)$ for $z<0$. Since the odd 
modes have a zero at the brane's position, their derivative is continuous at $z=0$ 
and, in this sense, they are unaffected by the brane. In the next subsections, 
whenever the  Schr\"odinger operators be invariant under $z\rightarrow -z$ we shall 
restrict the discussion to the even modes.

Summarizing, for the field fluctuations that lie along $\{\mathbf{T}_0\}$, 
we find that $\boldsymbol{\beta}$ does not propagates and the 
absence of modes with masses $m^2< 0$ for $\boldsymbol{\Xi}$ 
and $\boldsymbol{a}_\mu$ implies the perturbative stability of the domain 
wall configuration in this sector. Additionally, we find that the zero modes 
of $\boldsymbol{\Xi}$ and $\boldsymbol{a}_\mu$ do not generate localizable 4D 
massless modes on the core of the wall while their massive ones propagate in the 
bulk being scattered by the wall.

\subsection{Fluctuations along $\{\mathbf{T}_{\text{br}}\}$}

As shown in \cite{Pantoja:2015yin}, the $n_{\text{br}}$ field fluctuations 
$\boldsymbol{\varphi}\in\{\mathbf{T}_{\text{br}}\}$ would be 5D Nambu-Goldstone 
bosons if the symmetry were global rather than local and their 4D massless modes 
would be gravitationally trapped on the core of the wall. In the gauged model, all 
these scalar fluctuations can be gauged away by fixing 
$\boldsymbol{\varphi}=0$ (see Appendix \ref{appendixB}) and we are left with 
$n_{\text{br}}$ vector field fluctuations 
$\boldsymbol{\mathcal{A}}_a\in \{\mathbf{T}_{\text{br}}\}$, all with the same 5D 
\textit{mass} $\mathcal{M}_W$ (\ref{mass_w_1}) generated through a spontaneous 
gauge symmetry breaking.
 
Now, for the Lie algebra gauge invariant fluctuations $(\boldsymbol{\alpha}, \boldsymbol{\beta},\boldsymbol{a}_\mu)\in \{\mathbf{T}_{\text{br}}\}$, 
$\boldsymbol{\alpha}$ and $\boldsymbol{\beta}$ are not independent 
(\ref{alpha_beta_constraint}). We find
\begin{equation}\label{alpha_br}
\boldsymbol{\alpha}=i\texttt{g}(\mathcal{M}^{2}_{W})^{-1}[\boldsymbol{\Phi}^k,e^{-3A}\partial_z(e^{3A/2}\boldsymbol{\beta})],
\end{equation}
i.e. they correspond to a single physical perturbation, with the field fluctuation 
$\boldsymbol{\beta}$ satisfying
\begin{equation}
\partial^\mu\partial_\mu\boldsymbol{\beta} + \left(\frac{3}{2}A''-\frac{9}{4}A'^2-e^{2A}\mathcal{M}^2_{W}\right)\boldsymbol{\beta} +\partial_z^2\boldsymbol{\beta}=0,
\end{equation}
where $\mathcal{M}^2_W$ is given by (\ref{mass_w_1}). We see that there is a non-trivial mixing between the original fluctuations $\boldsymbol{\varphi}$, 
$\boldsymbol{\mathcal{A}}_z$ and $\boldsymbol{\chi}$.

\begin{figure}
\centerline{
    \includegraphics[width=0.3\textwidth,angle=0]{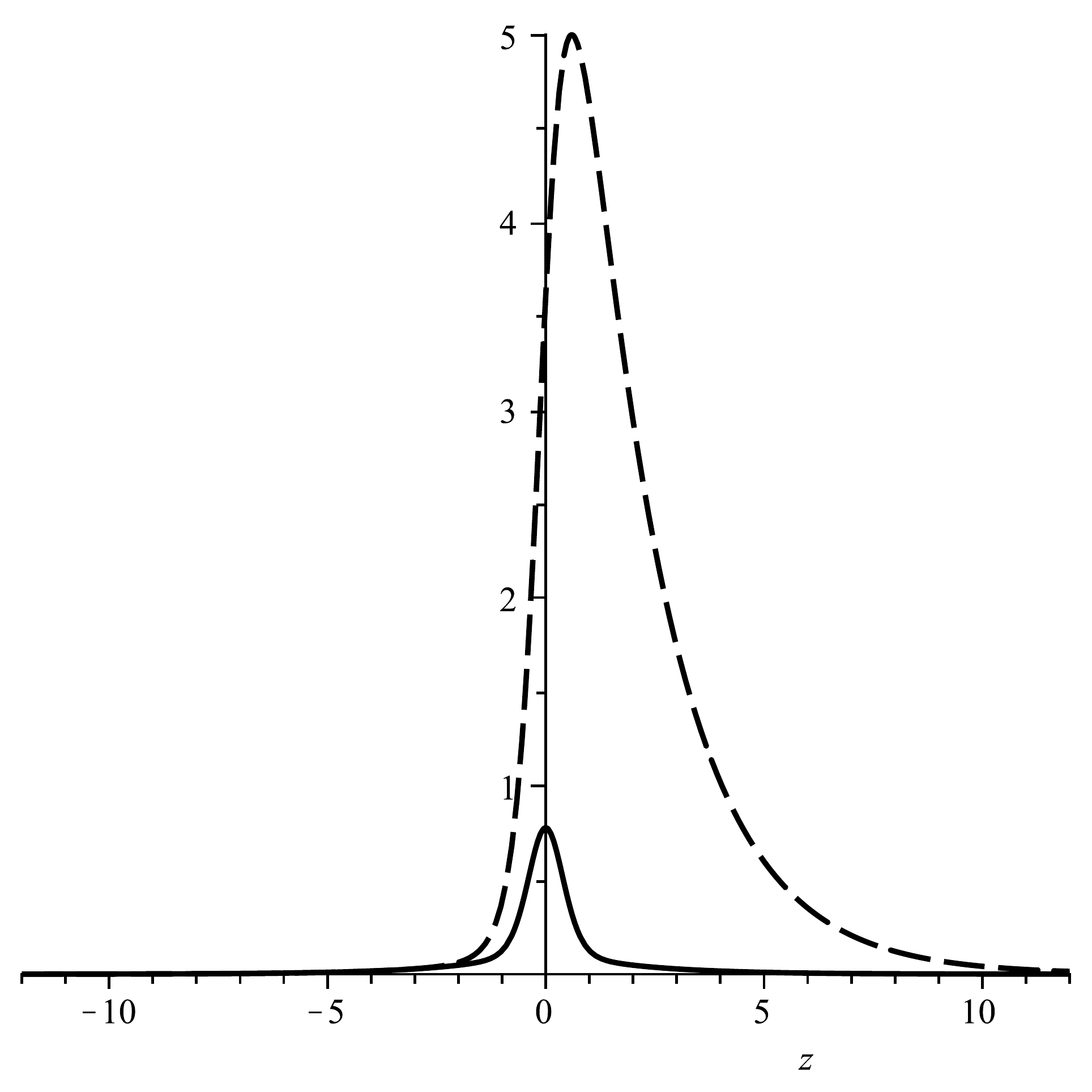}}
\caption{Schr\"odinger potentials $\mathcal{V}_2$ (solid line) and $\mathcal{V}_{4+}$ (dashed line) for the 
modes $\boldsymbol{\beta}\in\{\mathbf{T}_{\text{br}}\}$ and 
$\boldsymbol{\Omega}\in\{\mathbf{T}_+\}$, respectively. $\mathcal{V}_{4-}$ for the 
modes $\boldsymbol{\Omega}\in\{\mathbf{T}_-\}$ is the mirror image of 
$\mathcal{V}_{4+}$.}\label{shape_one}
\end{figure}

The modes $\boldsymbol{\beta}(x,z)\sim e^{ip\cdot x}\boldsymbol{\beta}(z)$ with $\boldsymbol{\beta}\in \{\mathbf{T}_{\text{br}}\}$,  satisfy 
the Schr\"odinger-like equation
\begin{equation}\label{A_y_br}
(-\partial_{z}^2 + \mathcal{V}_2)\boldsymbol{\beta}= m^2\boldsymbol{\beta},
\end{equation}
where 
\begin{equation}
\mathcal{V}_2= \frac{9}{4}A'^{2}-\frac{3}{2}A''  + e^{2A}\mathcal{M}^2_W.
\end{equation}
The Schr\"odinger operator in 
(\ref{A_y_br}) can be rewritten as
\[
(-\partial_{z}^2 + \mathcal{V}_2)=\left(\partial_z-\frac{3}{2}A'\right)\left(-\partial_z-\frac{3}{2}A'\right) +e^{2A}\mathcal{M}^2_W
\]
where the first term is a nonnegative definite operator on normalizable functions and the second term is never 
negative. Therefore, the eigenvalues of (\ref{A_y_br}) are always nonnegative and 
there are no tachyonic modes. In fact, $\mathcal{V}_2$ is everywhere positive with 
the shape of a symmetric potential barrier of finite height that vanishes asymptotically 
at $|z|\rightarrow\infty$ (see Fig.\ref{shape_one}). Hence, (\ref{A_y_br}) does not 
support bound states with $m^2\leq 0$, while those modes with $m^2>0$ behave as 
scattered waves by the wall. From these results and (\ref{alpha_br}) it follows that 
$\boldsymbol{\alpha}$ has no zero modes either.

Notice that the eventually large values of $\boldsymbol{\alpha}$ and 
$\boldsymbol{\beta}$ as $z\rightarrow\pm\infty$ mean that the perturbative theory is 
not trusted as we move far away from the core of the wall. Indeed, since the steps in 
the procedure to obtain (\ref{alpha_br}) and (\ref{A_y_br}) are purely formal, we must 
specify the behavior of fields at the $AdS_5$ boundary at $z=\pm\infty$ in order to 
pick out those solutions which are suitable for the description of the physical situation. 

\begin{figure}
\centerline{
    \includegraphics[width=0.3\textwidth,angle=0]{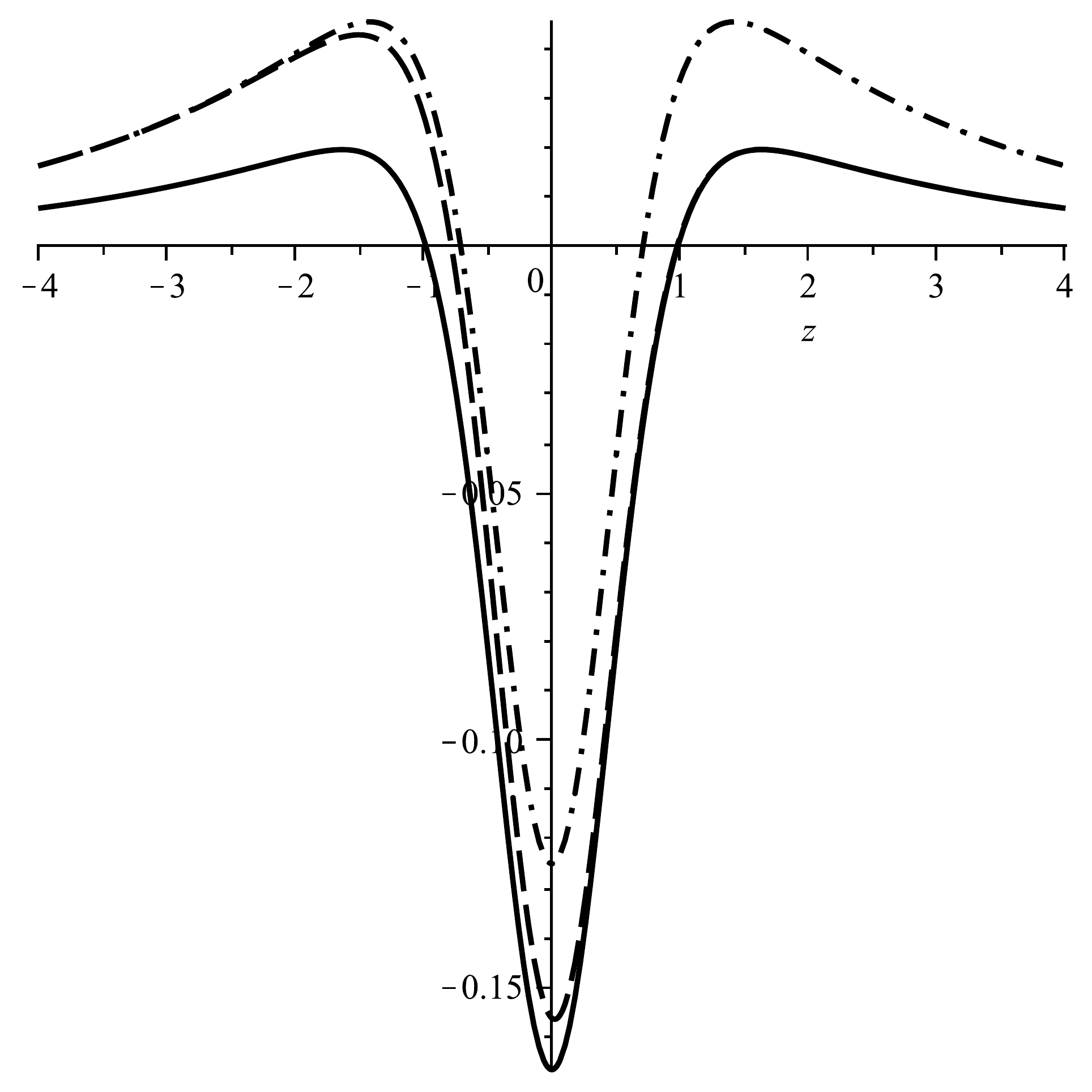}}
\caption{Schr\"odinger potentials $\mathcal{V}_1$ (solid line), $\mathcal{V}_3$ 
(dot-dashed line) and $\mathcal{V}_{5+}$ (dashed line) for the $\boldsymbol{a}_\mu$ 
modes along $\{\mathbf{T}_0\}$, $\{\mathbf{T}_{\text{br}}\}$ and $\{\mathbf{T}_+\}$, 
respectively. $\mathcal{V}_{5-}$ for the modes 
$\boldsymbol{a}_\mu\in\{\mathbf{T}_-\}$ is the mirror image of $\mathcal{V}_{5+}$.}
\label{shape_two}
\end{figure}

Next, let us consider the gauge invariant fluctuations $\boldsymbol{a}_\mu$. We find 
that the modes $\boldsymbol{a}_\mu(x,z)\sim e^{ip\cdot x}\boldsymbol{a}_\mu(z)$, 
satisfy 
\begin{equation}\label{gauge_QM}
(-\partial_{z}^2 + \mathcal{V}_3)\boldsymbol{a}_{\mu}= m^2\boldsymbol{a}_{\mu},
\end{equation}
where 
\begin{equation}
\mathcal{V}_3= \mathcal{V}_{1}+ e^{2A}\mathcal{M}_W^2,
\end{equation}
with $\mathcal{V}_{1}$ given by (\ref{VQ2}). The Schr\"odinger operator in (\ref{gauge_QM}) can be 
rewritten as
\[
(-\partial_{z}^2 + \mathcal{V}_3)=\left(\partial_z+\frac{1}{2}A'\right)\left(-\partial_z+\frac{1}{2}A'\right) +e^{2A}\mathcal{M}^2_W
\]
where the first term is a nonnegative definite operator and the second term is never 
negative. Therefore, the eigenvalues of (\ref{gauge_QM}) are nonnegative definite 
and there are no tachyonic modes. For $\texttt{g}^2<b^2/15$,  $\mathcal{V}_3$ has 
a volcano-like profile (see Fig.\ref{shape_two}) and one might naively expect that it 
supports a massless mode and a continuum of massive modes which propagates in 
the bulk. But, as is well known, a bulk mass term for a vector field does not allow for 
a 4D massless mode localized on the core of the wall 
\cite{Ghoroku:2001zu,Batell:2006dp}. Since $\mathcal{V}_3$ vanishes asymptotically 
at $|z|\rightarrow\infty$, there is no gap and the continuum modes have all possible 
$m^2>0$.

In order to examine the mode functions, let us suppose that we can approximate 
$A(z)$ by its brane limit (\ref{thin_wall_limit}).\footnote{\label{footnote2}We cannot 
take the brane limit without sending the scale of symmetry breaking $v$, and hence 
$\mathcal{M}_W^2$, to zero at the same time. However, the spectrum for $v\neq 0$ 
in the high curvature regime is expected to be qualitatively similar to the 
one obtained by approximating $A(z)$ as in (\ref{thin_wall_limit}) for 
$(\mathcal{M}_W/k)^2<<1<<1/v^2$, for $v$ not zero as long as $\texttt{g}$ is 
sufficiently small.} Imposing Neumann boundary conditions at $z=\pm\infty$ we find 
for the massive modes
\begin{equation}\label{massive_br_fluct}
(\boldsymbol{a}_\mu)_q\sim \xi^{1/2}\left[Y_\alpha(m\xi) + C_3J_\alpha(m\xi) \right]\varepsilon_\mu,
\end{equation}
where $\xi$ is given by (\ref{bessel_arg}), $\alpha=\sqrt{1+(\mathcal{M}_W/k)^2}$ 
and $C_3$ is a constant, determined by the continuity of $\boldsymbol{a}_\mu$ and 
the jump condition in $\partial_z \boldsymbol{a}_\mu$ at $z=0$, given by 
\begin{equation}
C_3=-\frac{Y_\alpha(m/k) + (m/k)Y'_\alpha(m/k)}{J_\alpha(m/k) +(m/k)J'_\alpha(m/k)}.
\end{equation}
There are no massless nor localized 4D massive modes.

Notice that the disappearance of the would be gravitationally trapped 4D 
Nambu-Goldstone modes along $\{\mathbf{T}_{\text{br}}\}$ does not 
generate localizable 4D massive modes for the gauge field fluctuations 
$\boldsymbol{a}_\mu\in\{\mathbf{T}_{\text{br}}\}$. On the other hand, if we set 
$\texttt{g}=0$, the coupling between $\boldsymbol{\alpha}$ and 
$\boldsymbol{\beta}$ disappears and $\boldsymbol{\beta}=0$. The 4D 
Nambu-Goldstone bosons $\in\{\mathbf{T}_{\text{br}}\}$ then reappear in the physical 
spectrum and the gauge bosons $\boldsymbol{a}_\mu \in \{\mathbf{T}_{\text{br}}\}$ 
behave as those that lie along $\{\mathbf{T}_0\}$. Obviously, from the 4D observers 
point of view the spectrum along $\{\mathbf{T}_{\text{br}}\}$ is discontinuous in the 
limit $\texttt{g}\rightarrow 0$. However, as it will be shown in the following, once the 
continuum modes are also taken into account, the spectrum is continuous 
in this limit. Discrete 4D massive vector field fluctuations along 
$\{\mathbf{T}_{\text{br}}\}$ appears, these being \textit{quasi-localized}.

In the brane approximation (\ref{thin_wall_limit}), (\ref{gauge_QM}) admits also a 
4D massive metastable 
mode with a complex eigenvalue $m^2=m_0^2-im_0\Gamma$, when radiative 
boundary conditions \cite{Balasubramanian:1999ri} at $z\rightarrow\pm\infty$ are 
imposed.\footnote{This 
effect is similar to that studied in Ref.\cite{Dubovsky:2000am} for a free 5D massive 
scalar field in the RS-brane background.} These metastable modes are given by 
\begin{equation}\label{resonance_br}
(\boldsymbol{a}_\mu)_q\sim \xi^{1/2}H^{(1)}_{\alpha}(m\xi)\varepsilon_\mu,
\end{equation}
where $H_{\alpha}^{(1)}=J_\alpha +iY_\alpha$ is the first Hankel function, while 
the continuity of $\boldsymbol{a}_\mu$ and 
the jump condition in $\partial_z \boldsymbol{a}_\mu$ at $z=0$ yield the eigenvalue condition
\[
\frac{m}{k}\frac{H_{\alpha-1}^{(1)}(m/k)}{H_\alpha^{(1)}(m/k)}= \alpha-1.
\] 
For $(\mathcal{M}_W/k)^2<<1$, with $(m_0/\mathcal{M}_W)^2<<1$ and 
$(\Gamma/m_0)^2<<1$, we find 
\begin{equation}\label{mass_resonance_br}
m_0^2=\frac{1}{2}\mathcal{M}^2_W\left(\frac{\mathcal{M}_W}{k}\right)^2,\quad \frac{\Gamma}{m_0}= \frac{\pi}{2}\left(\frac{\mathcal{M}_W}{k}\right)^2.
\end{equation}
It is seen from (\ref{resonance_br},\ref{mass_resonance_br}) that now a massive 
discrete mode with a finite lifetime exist, whose \textit{mass} and \textit{width} are 
suppressed by $\left({\mathcal{M}_W}/{k}\right)^2$. This mode decays into the 
continuum modes, due to its finite lifetime, and dissapears from the 
spectrum.       

To summarize, for the field fluctuations 
$(\boldsymbol{\alpha}, \boldsymbol{\beta},\boldsymbol{a}_\mu)\in\{\mathbf{T}_{\text{br}}\}$, 
we find no modes with $m^2<0$, i.e. the domain wall configuration is perturbatively 
stable in this sector. All these field fluctuations exhibit a tower of massive modes 
which propagate in the bulk, with a continuous spectrum for $m^2>0$ and no 
massless modes. On the other hand, as a consequence of the spontaneous gauge symmetry breaking, 
the gauge field excitations $\boldsymbol{a}_\mu$ become massive and get 
quasi-localizable discrete 4D massive modes that decay into the continuum modes 
due to their finite lifetime.

\subsection{Fluctuations along 
$\mathcal{K}^\bot=\{\mathbf{T}_+\}\oplus\{\mathbf{T}_-\}$}

To start with, it should be noted that if the $SU(5)$ symmetry were global rather than 
gauge, the scalars $\boldsymbol{\varphi}\in \mathcal{K}^\bot$ would be associated 
to fluctuations along the generators that are broken only at one side of the wall (see 
Eqs.(\ref{T_plus},\ref{T_minus})). In this case, gravitationally trapped 
4D Nambu-Goldstone fields appear corresponding to rotations of 
$\boldsymbol{\Phi}^k$ within the class described by $H_{\pm}/H_0$ 
\cite{Pantoja:2015yin}. On the other hand, in the gauged model we can not fix  
$\boldsymbol{\varphi}=0$ for $\boldsymbol{\varphi}\in \mathcal{K}^\bot$  (see 
Appendix \ref{appendixB}).
 
Now, for the gauge invariant fluctuations 
$(\boldsymbol{\alpha}, \boldsymbol{\beta},\boldsymbol{a}_\mu)\in \mathcal{K}^\bot$,  
$\boldsymbol{\alpha}$ and $\boldsymbol{\beta}$ satisfy the 
constraint
\begin{equation}\label{alpha_bot}
\boldsymbol{\alpha}=i\texttt{g} (\mathcal{M}^{2}_{\pm})^{-1}[\boldsymbol{\Phi}^k,e^{-3A}\partial_z(e^{3A/2}\boldsymbol{\beta})],
\end{equation}
where $\mathcal{M}^2_{\pm}$ is given by (\ref{mass_w_2}), with the plus sign for 
$(\boldsymbol{\alpha}, \boldsymbol{\beta})\in \{\mathbf{T}_+\}$ and the minus sign for 
$(\boldsymbol{\alpha}, \boldsymbol{\beta})\in \{\mathbf{T}_-\}$. 
Hence, $\boldsymbol{\alpha}$ and $\boldsymbol{\beta}$ are not 
independent and correspond to a single physical perturbation. 
The gauge invariant field fluctuation $\boldsymbol{\beta}$ satisfy
\begin{eqnarray}\label{beta_bot}
&&\partial^\mu\partial_\mu\boldsymbol{\beta} + \left(\frac{3}{2}A''-\frac{9}{4}A'^2-e^{2A}\mathcal{M}^2_{\pm}\right)\boldsymbol{\beta} + \partial_z^2\boldsymbol{\beta}\nonumber\\&&-\left(\partial_z\ln\mathcal{M}^2_{\pm}\right)\left(\frac{3}{2}A' \boldsymbol{\beta} + \partial_z\boldsymbol{\beta})\right)=0.
\end{eqnarray} 
As in the previous sector, this implies a non-trivial mixing between the original 
fluctuations $\boldsymbol{\varphi}$, $\boldsymbol{\mathcal{A}}_z$ and 
$\boldsymbol{\chi}$, this time with the background playing a prominent role. 

Let $\boldsymbol{\Omega}$ be defined as 
\begin{equation}
\boldsymbol{\Omega}\equiv (\mathcal{M}_{\pm})^{-1}\boldsymbol{\beta}.
\end{equation}
The modes $\boldsymbol{\Omega}(x,z)\sim e^{ip\cdot x}\boldsymbol{\Omega}(z)$ with $\boldsymbol{\Omega}\in \{\mathbf{T}_{\text{br}}\}$,  satisfy 
the Schr\"odinger-like equation
\begin{equation}\label{beta_QM_bot}
(-\partial_{z}^2 + \mathcal{V}_{4\pm})\boldsymbol{\Omega}= m^2\boldsymbol{\Omega},
\end{equation}
where 
\begin{eqnarray}
\mathcal{V}_{4\pm}&&= \frac{9}{4}A'^{2}-\frac{3}{2}A''  + e^{2A}\mathcal{M}^2_{\pm} + \frac{1}{4}\left((\ln \mathcal{M}^2_{\pm})'\right)^2 \nonumber\\&&- \frac{1}{2}(\ln \mathcal{M}^2_{\pm})'' + \frac{3}{2}A' (\ln \mathcal{M}^2_{\pm})',
\end{eqnarray}
with $\mathcal{V}_{4\pm}\rightarrow\mathcal{V}_{4\mp}$ under $z\rightarrow -z$.
The Schr\"odinger operator in (\ref{beta_QM_bot}) can be 
rewritten as
\begin{equation}\label{susy_beta_QM_bot}
(-\partial_{z}^2 + \mathcal{V}_{4\pm})=\left(\partial_z-\frac{3}{2}Q'_\pm\right)\left(-\partial_z-\frac{3}{2}Q'_\pm\right) +e^{2A}\mathcal{M}^2_{\pm},
\end{equation}
where
\begin{equation}
Q_{\pm}\equiv A +\frac{1}{3}\ln \mathcal{M}^2_{\pm}.
\end{equation}
It follows that the eigenvalues of (\ref{beta_QM_bot}) 
are nonnegative definite and there are no tachyonic modes. In fact, 
(\ref{beta_QM_bot}) is 
just a Schr\"odinger equation with an asymmetric potential barrier of finite height 
that vanishes asymptotically at $|z|\rightarrow\infty$ (see Fig.\ref{shape_one}). 
Hence, (\ref{beta_QM_bot}) does not support bound states with $m^2\leq 0$, while 
those modes with $m^2>0$ behave as scattered waves by the wall. From these 
results and (\ref{alpha_bot}) it follows that $\boldsymbol{\alpha}$ has no tachyonic 
nor normalizable zero modes neither.

Finally, the modes $\boldsymbol{a}_\mu(x,z)\sim e^{ip\cdot x}\boldsymbol{a}_\mu(z)$ 
of the gauge invariant fluctuation $\boldsymbol{a}_\mu$, satisfy 
\begin{equation}\label{gauge_QM_bot}
(-\partial_{z}^2 + \mathcal{V}_{5\pm})\boldsymbol{a}_{\mu}= m^2\boldsymbol{a}_{\mu},
\end{equation}
where 
\begin{equation}\label{gauge_potential_bot}
\mathcal{V}_{5\pm}= \mathcal{V}_{1}+ e^{2A}\mathcal{M}_{\pm}^2,
\end{equation}
with $\mathcal{V}_{1}$ given by (\ref{VQ2}) and $\mathcal{V}_{5\pm}\rightarrow\mathcal{V}_{5\mp}$ under $z\rightarrow -z$. The Schr\"odinger operator in 
(\ref{gauge_QM_bot}) can be rewritten as
\[
(-\partial_{z}^2 + \mathcal{V}_{5\pm})=\left(\partial_z+\frac{1}{2}A'\right)\left(-\partial_z+\frac{1}{2}A'\right) +e^{2A}\mathcal{M}^2_{\pm}.
\]
Therefore, in the spectrum of (\ref{gauge_QM_bot}) there are no negative 
eigenvalues and we have no tachyonic modes. For $\texttt{g}^2<4b^2/15$, 
the potential has an asymmetric volcano-like profile that vanishes asymptotically at 
$|z|\rightarrow\infty$ (see Fig.\ref{shape_two}). Hence, as in the previous Lie 
algebra sector, it supports no localizable massless modes and a continuum of 
massive modes with all the possible $m^2>0$. Obviously there are no localizable 4D massive modes neither. 

The absence of the would be gravitationally trapped 4D Nambu-Goldstone fields 
if the symmetry were global than local \cite{Pantoja:2015yin} and of localizable 4D 
massive modes for the gauge field, suggest the existence of quasi-localizable 4D 
massive modes for the last one, in order to make continuous the zero gauge 
coupling limit of the spectrum in this Lie algebra sector of the fluctuations. 

To determine the existence of metastable states 
$\boldsymbol{a}_\mu\in \mathcal{K}^\bot$, we approximate 
$A(z)$ as in (\ref{thin_wall_limit}) and $\mathcal{M}^2_{\pm}$ 
by\footnote{See footnote \ref{footnote2}.}
\begin{equation}\label{brane_masses_Kbot}
\mathcal{M}^2_+\sim \begin{cases}
                                       \quad 0, \quad \,\,\,z>0\\
                                       \mathcal{M}^2_W,\quad \!z<0
                                       \end{cases}\!\!,\quad
\mathcal{M}^2_-\sim \begin{cases}
                                       \mathcal{M}^2_W,\quad \!z<0\\
                                       \quad 0, \quad \,\,\,z>0
                                       \end{cases}\!\!.                                        
\end{equation}
Now, for the radiative boundary problem, $\boldsymbol{a}_\mu(z)\in\{\mathbf{T}_+\}$ 
is given by
\begin{equation}\label{massive_+_fluct}
(\boldsymbol{a}_\mu)_q\sim \varepsilon_\mu\xi^{1/2}\begin{cases}
                                            C_4 H_1^{(1)}(m\xi),\quad z>0\\
                                            C_5H^{(1)}_\alpha(m\xi),\quad z<0
                                            \end{cases}
\end{equation}
where $\xi$ is given by (\ref{bessel_arg}),  $\alpha=\sqrt{1+(\mathcal{M}_W/k)^2}$ 
and the coefficients $C_4$ and $C_5$ determined, as before, by imposing the 
continuity of $\boldsymbol{a}_\mu(z)$ and the discontinuity of its first derivative at 
$z=0$, which must be $-k\boldsymbol{a}_\mu(0)$. The latter condition leads to the eigenvalue formula
\[
\frac{m}{k}\left[\frac{H_{\alpha-1}^{(1)}(m/k)}{H_\alpha^{(1)}(m/k)} + \frac{H_0^{(1)}(m/k)}{H_1^{(1)}(m/k)}\right]= \alpha-1.
\]  
For $(\mathcal{M}_W/k)^2<<1$, we find the same resonance massive given by 
(\ref{mass_resonance_br}). The quasi-localized 4D massive modes 
$\boldsymbol{a}_\mu\in\{\mathbf{T}_-\}$ can be obtained from (\ref{massive_+_fluct}) 
under $z\rightarrow -z$. 

The absence of tachyonic modes for the field excitations 
$(\boldsymbol{\alpha}, \boldsymbol{\beta}, \boldsymbol{a}_\mu)\in\mathcal{K}^\bot$, 
means that the domain wall configuration is perturbatively stable also in this sector. 
All these fluctuations exhibit a tower of 4D massive modes which propagate in the 
bulk, with a continuous spectrum for $m^2>0$ and no localized 4D massless modes.
As in the Lie algebra sector $\{\mathbf{T}_{\text{br}}\}$, we find metastable 4D 
massive gauge fluctuations in this sector also.

\section{Summary and Conclusions}\label{conclusions}

In terms of diffeomorphism- and Lie algebra gauge- invariant excitations, we have 
proven the perturbative stability of some topologically non-trivial 5D selfgravitating 
$SU(5)\times Z_2$ domain wall configurations. As expected, gravitational tensor and 
vector fluctuations, which are unchanged under Lie algebra gauge transformations, 
behave like its counterparts in the standard $Z_2$ domain walls. 

The behavior of the Lie algebra valued fluctuations is, of course, much more 
interesting. All exhibit towers of 4D massive modes which propagate in the bulk, 
with a continuous spectrum for $m^2 > 0$. All the would be 4D Nambu-Goldstone 
excitations associated to the partial breaking $SU(5)\times Z_2 \rightarrow H_0$ 
(gravitationally trapped if the SU(5) symmetry were global \cite{Pantoja:2015yin} 
rather than gauge) disappear from the physical 4D spectrum. No 4D massless gauge 
field excitations are found, and the massive ones are not localized. As we have seen, 
discrete metastable 4D massive mode functions for the gauge field fluctuations exist 
along the Lie algebra sectors where the 4D Nambu-Goldstone fields appear. Thus, 
an interesting version of the Higgs phenomenon takes place in these systems, 
whereby 4D gauge fluctuations along the spontaneously broken gauge sectors 
acquire masses and then escape from the core of the wall into the bulk.

Selfgravitating Higgs domain walls, of the sort studied here, provide perturbatively 
stable minimal settings with enhanced symmetry breaking patterns. Depending of the
nature of this pattern, these backgrounds are ideal to discuss \cite{Melfo:2011ev} the 
Dvali-Shifman mechanism of gauge field localization via bulk confinement 
\cite{Dvali:1996xe}. Indeed, if one wishes to construct a phenomenological viable 
non-abelian domain wall braneworld, the explicit gauge group $H_0$ on the core of 
the wall should be more akin to the standard model group. We hope to return to this 
and other related issues in the near future.
     
\section{Acknowledgements}
The author is deeply indebted to Alejandra Melfo for inspiring discussions and 
valuable comments. The author wishes also to thank Alba Ramirez for enjoyable collaboration.

\appendix
\section{Linearized equations for the fluctuations}\label{appendixA}

Following the procedure outlined in section \ref{fluctuations}, it is straightforward to 
derive perturbative equations to first order for fluctuations 
$(\boldsymbol{\varphi}, \boldsymbol{\mathcal{A}}_a, h_{ab})$ around a solution 
$({\boldsymbol{\Phi}}, {\boldsymbol{A}}_a;g_{ab})$ of the field 
equations (\ref{field_eq1},\ref{field_eq2},\ref{field_eq3},\ref{field_eq4}). We find
\begin{eqnarray}\label{fluctuations1}
&-&\frac{1}{2}g^{cd}\nabla_{c}\nabla_{d}h_{ab}+R^{c}{}_{(ab)}{}^{d}h_{cd}+R^{c}_{(a}h^{ }_{b)c}+\nabla_{(a}\nabla^{c}h_{b)c}\nonumber\\
&-&\frac{1}{2}\nabla_{a}\nabla_{b}\left(g^{cd}h_{cd}\right)=4\text{Tr}\{\mathbf{D}_{(a}\boldsymbol{\Phi}\mathbf{D}_{b)}\boldsymbol{\varphi}\}+\frac{2}{3}h_{ab}V(\boldsymbol{\Phi})\nonumber\\&+&\frac{2}{3}\!\!\left(\frac{\partial V(\boldsymbol{\Phi})}{\partial \phi_q}\varphi_q \right) \!\!g_{ab}
+ 4{i}{\texttt{g}}\text{Tr}\{\mathbf{D}_{(a}\boldsymbol{\Phi}[\boldsymbol{\mathcal{A}}_{b)},\boldsymbol{\Phi}]\}\nonumber\\ &+&2h_{cd}\text{Tr}\{\mathbf{F}_{a}^{\,\,c}\mathbf{F}_{b}^{\,\,d}\}
+4g^{cd}\text{Tr}\{\mathbf{F}_{ac}\mathbf{D}_{[b}\boldsymbol{\mathcal{A}}_{d]})
+\mathbf{F}_{bc}\mathbf{D}_{[a}\boldsymbol{\mathcal{A}}_{d]})\}\nonumber\\
&-&\frac{1}{3}h_{ab}\text{Tr}\{\mathbf{F}_{cd}\mathbf{F}^{cd}\}
-\frac{2}{3}g_{ab}\text{Tr}\{\mathbf{F}^{cd}\mathbf{D}_{[c}\boldsymbol{\mathcal{A}}_{d]}\},
\end{eqnarray}
where $R^{d}{}_{abc}$ and $R^b_{\,a}$ are the Riemann and Ricci curvatures of $g_{ab}$, respectively,
\begin{eqnarray}\label{fluctuations2}
&-&\frac{1}{2}g^{ab}g^{cd}\left(\nabla_{a}h_{bd}+\nabla_{b}h_{ad}-\nabla_{d}h_{ab}\right)\mathbf{D}_{c}\boldsymbol{\Phi} \nonumber\\
&+&ig^{ab}\texttt{g}\mathbf{D}_b[\boldsymbol{\mathcal{A}}_a,\boldsymbol{\Phi}]+ig^{ab}\texttt{g}[\boldsymbol{\mathcal{A}}_a,\mathbf{D}_b\boldsymbol{\Phi}]\nonumber\\&-&h^{ab}\mathbf{D}_{a}\mathbf{D}_{b}{\boldsymbol{\Phi}}+g^{ab}\mathbf{D}_{a}\mathbf{D}_{b}\boldsymbol{\varphi}=\frac{\partial^2 V(\boldsymbol{\Phi})}{\partial\phi_p\partial\phi_q}\varphi_p\mathbf{T}^q,
\end{eqnarray}
where
\begin{equation}\label{delta_V_2}
V(\boldsymbol{\Phi} +\boldsymbol{\varphi})=V(\boldsymbol{\Phi}) + \frac{\partial V(\boldsymbol{\Phi})}{\partial\phi_q}\varphi_q + \frac{1}{2}\frac{\partial^2 V(\boldsymbol{\Phi})}{\partial\phi_p\partial\phi_q}\varphi_p\varphi_q + O(\boldsymbol{\varphi}^3)
\end{equation}
and 
\begin{eqnarray}\label{fluctuations3}
&&g^{ac}\mathbf{D}_c\boldsymbol{\mathcal{F}}_{ab} + i\texttt{g}g^{ac}[\boldsymbol{\mathcal{A}}_c,\mathbf{F}_{ab}] +\texttt{g}^2[\boldsymbol{\Phi}, [\boldsymbol{\mathcal{A}}_b,\boldsymbol{\Phi}]]\nonumber\\
&&-\frac{1}{2}g^{ac}g^{de}\left( (\nabla_ch_{be} + \nabla_b h_{de} - \nabla_e h_{cb})\mathbf{F}_{ad}\right.\nonumber \\&&\left.-  (\nabla_ch_{ae} + \nabla_a h_{de} - \nabla_e h_{ca})\mathbf{F}_{bd}\right)-h^{ac}D_c\mathbf{F}_{ab}\nonumber\\
&&-i\texttt{g}[\boldsymbol{\varphi},\mathbf{D}_b\boldsymbol{\Phi}] - i\texttt{g}[\boldsymbol{\Phi},\mathbf{D}_b\boldsymbol{\varphi}] =0,
\end{eqnarray}
where $\boldsymbol{\mathcal{F}}_{ab}\equiv \mathbf{D}_a\boldsymbol{\mathcal{A}}_b-\mathbf{D}_b\boldsymbol{\mathcal{A}}_a$.
Clearly, (\ref{fluctuations1},\ref{fluctuations2},\ref{fluctuations3}) are Lie algebra 
gauge covariant with respect to the general background 
$({\boldsymbol{\Phi}}, {\boldsymbol{A}}_a;g_{ab})$, since $\boldsymbol{A}_a$ appears only in 
$\mathbf{F}_{ab}$ and in the covariant derivative $\mathbf{D}_a$.

Next, within the gauge-equivalent classes of background domain wall configurations 
$(\tilde{\boldsymbol{\Phi}}^k, \tilde{\boldsymbol{A}}^k_a;g^k_{ab})$, for simplicity we write (\ref{fluctuations1},\ref{fluctuations2},\ref{fluctuations3}) 
in the (Lie algebra) gauge fixed domain wall background 
$(\boldsymbol{\Phi}^k,\mathbf{0}_a;g_{ab}^k)$. We find
\begin{eqnarray}\label{fluctuations1_DW}
&-&\frac{1}{2}g^{cd}\nabla_{c}\nabla_{d}h_{ab}+R^{c}{}_{(ab)}{}^{d}h_{cd}+R^{c}_{(a}h^{ }_{b)c}+\nabla_{(a}\nabla^{c}h_{b)c}\nonumber\\
&-&\frac{1}{2}\nabla_{a}\nabla_{b}\left(g^{cd}h_{cd}\right)=4\text{Tr}\{\nabla_{(a}\boldsymbol{\Phi}^k\nabla_{b)}\boldsymbol{\varphi}\}+\frac{2}{3}h_{ab}V(\boldsymbol{\Phi}^k)\nonumber\\&+&\frac{2}{3}\!\!\left.\frac{\partial V(\boldsymbol{\Phi})}{\partial \phi_q}\right|_{\boldsymbol{\Phi}^k} \!\varphi_qg_{ab},
\end{eqnarray}
where now $R^{d}{}_{abc}$ and $R^b_{\,a}$ are the Riemann and Ricci curvatures of $g^k_{ab}$, respectively,
\begin{eqnarray}\label{fluctuations2_DW}
&&-\frac{1}{2}g^{ab}g^{cd}\left(\nabla_{a}h_{bd}+\nabla_{b}h_{ad}-\nabla_{d}h_{ab}\right)\nabla_{c}\boldsymbol{\Phi}^k \nonumber\\
&&+ig^{ab}\texttt{g}\nabla_b[\boldsymbol{\mathcal{A}}_a,\boldsymbol{\Phi}^k]+ig^{ab}\texttt{g}[\boldsymbol{\mathcal{A}}_a,\nabla_b\boldsymbol{\Phi}^k]\nonumber\\&&-h^{ab}\nabla_{a}\nabla_{b}{\boldsymbol{\Phi}^k}+g^{ab}\nabla_{a}\nabla_{b}\boldsymbol{\varphi}=\varphi_p\!\left.\frac{\partial^2 V(\boldsymbol{\Phi})}{\partial\phi_p\partial\phi_q}\right|_{\boldsymbol{\Phi}^k}\!\!\mathbf{T}^q
\end{eqnarray}
and
\begin{eqnarray}\label{fluctuations3_DW}
\!\!\!g^{ac}\nabla_c\left(\nabla_a\boldsymbol{\mathcal{A}}_b-\nabla_b\boldsymbol{\mathcal{A}}_a\right)&& \,+\,\texttt{g}^2[\boldsymbol{\Phi}^k, [\boldsymbol{\mathcal{A}}_b,\boldsymbol{\Phi}^k]]=\nonumber\\
+i\texttt{g}&&[\boldsymbol{\varphi},\nabla_b\boldsymbol{\Phi}^k] + i\texttt{g}[\boldsymbol{\Phi}^k,\nabla_b\boldsymbol{\varphi}],
\end{eqnarray}
where
\begin{equation}
\left.\frac{\partial V(\boldsymbol{\Phi})}{\partial\phi_q}\right|_{\boldsymbol{\Phi}^k}= \left(\phi_M'' + 4A'\phi_M'\right)\delta^{Mq}
\end{equation}
and the hessian of $V(\boldsymbol{\Phi})$ at $\boldsymbol{\Phi}^k$, $\left.{\partial^2 V(\boldsymbol{\Phi})}/{\partial\phi_p\partial\phi_q}\right|_{\boldsymbol{\Phi}^k}$, is a 
block-diagonal $(5^2-1)\times(5^2-1)$ matrix. 

In obtaining (\ref{fluctuations1_DW}) we have used
\[
\text{Tr}\{\partial_{(a}\boldsymbol{\Phi}^k[\boldsymbol{\mathcal{A}}_{b)},\boldsymbol{\Phi}^k]\}=\delta^y_{(a}\text{Tr}\{\boldsymbol{\mathcal{A}}^{\,}_{b)}[\boldsymbol{\Phi}^k(y),\partial_y\boldsymbol{\Phi}^k(y)]\}=0
\]
because $[\boldsymbol{\Phi}^k(y),\partial_y\boldsymbol{\Phi}^k(y)]=0$. Additionally, 
if we take the divergence of (\ref{fluctuations3_DW}), we find an integrability condition 
that can be used to remove some of the degrees of freedom.

Finally, (\ref{fluctuations3_DW}) is rewritten as
\begin{eqnarray}\label{fluctuations3_DW.2}
\frac{1}{2}\delta^{qp}g^{ac}\nabla_c(&&\nabla_a\boldsymbol{\mathcal{A}}_b-\nabla_b\boldsymbol{\mathcal{A}}_a)_p-\frac{1}{2}({M}^2)^{qp}(\boldsymbol{\mathcal{A}}_b)_p=\nonumber\\&&i\texttt{g}\text{Tr}\left\{\boldsymbol{\varphi}[\partial_b\boldsymbol{\Phi}^k,\mathbf{T}^q]+\partial_b\boldsymbol{\varphi}[\mathbf{T}^q,\boldsymbol{\Phi}^k]\right\},
\end{eqnarray}
where
\begin{equation}
({M}^2)^{qp}\equiv -2\texttt{g}^2\text{Tr}\left\{[\mathbf{T}^q,\boldsymbol{\Phi}^k(y)][\mathbf{T}^p,\boldsymbol{\Phi}^k(y)]\right\}.
\end{equation}

\section{On the Lie algebra gauge fixing for $(\boldsymbol{\varphi}, \boldsymbol{\mathcal{A}}_a)\notin \{\mathbf{T}_0\}$}\label{appendixB}

Here we show explicitly some issues related to the Lie algebra gauge fixing for 
$(\boldsymbol{\varphi}, \boldsymbol{\mathcal{A}}_a)\notin \{\mathbf{T}_0\}$ in the 
symmetry breaking {\bf A} (the symmetry breaking {\bf B} differing only in numerical 
factors). 

Under infinitesimal gauge transformations, we find
\begin{equation}
\varphi_{q}\mapsto \varphi_{q} + v\sqrt{\frac{5}{2}}\sigma_{q'},
\end{equation}
where $\mathbf{T}^q,\mathbf{T}^{q'} \in \{\mathbf{T}_{\text{br}}\}$. For 
$\varphi_{q}$ bounded, it follows that 
we can choose $\sigma_{q'}$ in order to make $\varphi_{q}=0$ whenever 
$\mathbf{T}^q \in \{\mathbf{T}_{\text{br}}\}$. Hence, using these $n_{\text{br}}$ 
gauge degrees of freedom, we can fix $\boldsymbol{\varphi}=0$ for 
$\boldsymbol{\varphi}\in \{\mathbf{T}_{\text{br}}\}$.

On the other hand, we find
\begin{equation}
\varphi_{q}\mapsto \varphi_{q} +v\frac{1}{2}\sqrt{\frac{5}{2}}(F+1)\sigma_{q'},
\end{equation}
where $\mathbf{T}^q,\mathbf{T}^{q'} \in \{\mathbf{T}_-\}$ and
\begin{equation}
\varphi_{p}\mapsto \varphi_{p} +v\frac{1}{2}\sqrt{\frac{5}{2}}(F-1)\sigma_{p'},
\end{equation}
where $\mathbf{T}^p,\mathbf{T}^{p'} \in \{\mathbf{T}_+\}$, with $F$ given by 
(\ref{F}). If $\varphi_{q}$ [$\varphi_{p}$] is a bounded function that decays to zero  
faster than $\sim v(F+1)$ [$\sim v(F-1)$] as 
$y\rightarrow-\infty$ [$y\rightarrow+\infty$], we could choose 
$\sigma_{q'}$ [$\sigma_{p'}$] 
to gauge away $\varphi_{q}$ [$\varphi_{p}$]. But for fluctuations $\varphi_{q}$ 
[$\varphi_{p}$] that do not vanish away of the core of the wall, it 
is clear that this will require a growing $\sigma_{q'}$ [$\sigma_{p'}$] as 
$y\rightarrow-\infty$ [$y\rightarrow+\infty$], which conflicts with $\sigma_{q'}$ 
[$\sigma_{p'}$] being small. It follows that, for general bounded fluctuations 
$(\boldsymbol{\varphi}, \boldsymbol{\mathcal{A}}_a)\in \mathcal{K}^\bot$, we cannot 
fix to zero the 5D scalar fluctuation $\boldsymbol{\varphi}$ all along the additional 
dimension.

\end{document}